\documentclass[%
 aip,
 amsmath,amssymb,
 reprint,%
]{revtex4-1}

\usepackage{graphicx}% Include figure files
\usepackage{dcolumn}% Align table columns on decimal point
\usepackage{bm}% bold math
\usepackage{caption} % For caption spacing
\usepackage{subcaption} % For sub-figures
\usepackage{xr}
\usepackage[utf8]{inputenc}
\usepackage[T1]{fontenc}
\usepackage{mathptmx}
\usepackage{todonotes}
\usepackage{natbib}
\begin{document}

\preprint{AIP/123-QED}

\title[Abundance of cavity-free polaritonic states in resonant materials and nanostructures]{Abundance of cavity-free polaritonic states in resonant materials and nanostructures}
% Force line breaks with \\

\author{Adriana Canales}
\affiliation{Department of Physics, Chalmers University of Technology, 412 96, Göteborg, Sweden.
}%

\author{Denis G. Baranov}%
\affiliation{Department of Physics, Chalmers University of Technology, 412 96, Göteborg, Sweden.
}%
\affiliation{Center for Photonics and 2D Materials, Moscow Institute of Physics and Technology, Dolgoprudny 141700, Russia.
}

\author{Tomasz J. Antosiewicz}
\affiliation{Faculty of Physics, University of Warsaw, Pasteura 5, 02-093, Warsaw, Poland.
}%
\affiliation{Department of Physics, Chalmers University of Technology, 412 96, Göteborg, Sweden.
}

\author{Timur Shegai}%
 \email{timurs@chalmers.se.}
\affiliation{Department of Physics, Chalmers University of Technology, 412 96, Göteborg, Sweden.
}%

\date{\today}% It is always \today, today,

\begin{abstract}
Strong coupling between various kinds of material excitations and optical modes has recently shown potential to modify chemical reaction rates in both excited and ground states. The ground-state modification in chemical reaction rates has usually been reported by coupling a vibrational mode of an organic molecule to the vacuum field of an external optical cavity, such as a planar Fabry-Pérot microcavity made of two metallic mirrors. However, using an external cavity to form polaritonic states might: (i) limit the scope of possible applications of such systems, and (ii) be unnecessary. Here we highlight the possibility of using optical modes sustained by materials themselves to \emph{self-couple} to their own electronic or vibrational resonances. By tracing the roots of the corresponding dispersion relations in the complex frequency plane, we show that electronic and vibrational polaritons are natural eigenstates of bulk and nanostructured resonant materials that require no external cavity. Several concrete examples, such as a slab of excitonic material and a spherical water droplet in vacuum are shown to reach the regime of such cavity-free self-strong coupling. The abundance of cavity-free polaritons in simple and natural structures questions their relevance and potential practical importance for the emerging field of polaritonic chemistry, exciton transport, and modified material properties.   

\end{abstract}

\maketitle

%%%%%%%%%%%%%%%%%%%%%%%%%%%%%    
%%%%INTRODUCTION     
%%%%%%%%%%%%%%%%%%%%%%%%%%%% 
\section{Introduction}

Strong coupling is a distinct regime of light-matter interactions, which is reached when resonant optical modes and material excitations (electronic, vibrational, etc.) exchange energy faster than they lose it to the environment. This fast energy exchange gives rise to new quasiparticles;  \emph{polaritons}. \cite{torma2014strong,Baranov2018}

Following the achievement of strong coupling, fundamental questions have arisen, such as whether polaritonic states could influence material and chemical properties.\cite{ebbesen2016hybrid,feist2018polaritonic} The possibility of affecting chemical reactions in the excited state (i.e. photo-chemistry) seems to be quite well understood today. Several experimental studies show that strong coupling can affect photo-chemical processes \cite{hutchison2012modifying,Munkhbat2018,Peters:19} and these results are essentially in agreement with theoretical descriptions.\cite{PhysRevLett.119.136001, galego2016suppressing, PhysRevLett.116.238301, ribeiro2018polariton, fregoni2018manipulating, felicetti2020photoprotecting, nefedkin2020role} Thermally-activated ground-state chemical reactions in the vibrational strong coupling (VSC) regime seem to be more controversial. In this case, just a few experimental reports exist \cite{thomas2016ground,Thomas2019,lather2019cavity,hirai2020modulation} and, there is far less conclusive agreement with theory.\cite{flick2017atoms, martinez2018can, galego2019cavity, campos2019resonant, schafer2019modification, campos2020polaritonic, zhdanov2020vacuum, li2020cavity} 

An example of this controversy is given by common intuition, by stating that ground-state chemical reactions are local, i.e. the chemistry at position A does not depend on any far-away position B. Nevertheless, recent experimental studies  \cite{thomas2016ground,Thomas2019,lather2019cavity,hirai2020modulation} suggest that such a local approximation to the chemistry might not always hold. Contemporary empirical observations,\cite{ebbesen2016hybrid, feist2018polaritonic, yuen2019polariton} indicate that the chemistry can be affected by resonant, collective and non-local variables. For example, the reaction probability at a specific position depends on the far away presence (or absence) of metallic mirrors forming a Fabry-Pérot (FP) cavity. Specifically, it depends on the set of resonant optical modes that the mirrors form. This set of resonant optical modes and its relevance to chemistry is especially important, since they distinguish polaritonic chemistry from other common non-local but non-resonant contributions, such as screening, electrostatic, or solvent polarity effects.\cite{peters2015control, lee2017charge, peters2019control} Moreover, while standard chemical theory states that the reaction rate depends on the concentration of molecules, concentration-dependent collective Rabi splitting influences the rate further by modifying the energy levels. Although this train of thought is heavily debated in the community, mainly due to a lack of theoretical understanding of the recent experimental observations, we believe the potential of this new research frontier is quite high. Of course, further experiments and theoretical developments are needed to clarify this and other subtle issues.

%% Remark: Nonlocality is usually observed in wave mechanics, related to light waves or to quantum mechanical probability waves. It is a consequence of corresponding Maxwell and Schrödinger equations. However, although molecules of course obey quantum mechanics and thus are "wavy" on a molecular scale, they behave in a much more rigid way when it comes to ground-state chemistry, which to the first approximation should not be governed by Maxwell equations, and thus be local in this sense.
%% An example of nonlocality in chemistry though is chemistry close to metal surfaces. Since metals are highly polarizable they will always form image charges of the molecules. This will cause screening and hence modification in chemical properties. This type of nonlocality is however nonresonant. Effectively, this type of process is quite similar to the well-known "solvent effect" in chemistry. Solvents of different polarities have different dielectric functions (and thus screening), which can affect chemistry. Metals affects chemistry is a similar effective "solvent effect".
Without trying to resolve the discrepancy of the ground-state chemistry in the VSC regime at this point, here we wish to focus on another potentially relevant issue. The so far experimentally reported observations of chemistry modifications under VSC exclusively use FP cavities comprised of two metallic mirrors.\cite{thomas2016ground, Thomas2019, lather2019cavity, hirai2020modulation} Then, a natural question is if the role of the mirrors in these experiments is only to confine electromagnetic modes in order to form polaritons, or if it goes beyond that. Thus the importance of discussing the concept of ``cavity'' in the strong coupling regime and understanding its relevance to polaritonic chemistry. 

\begin{figure}[t] %%%%%%%%%%FIG.1%%%%%%%%
  \centering
    \includegraphics[width=\linewidth]{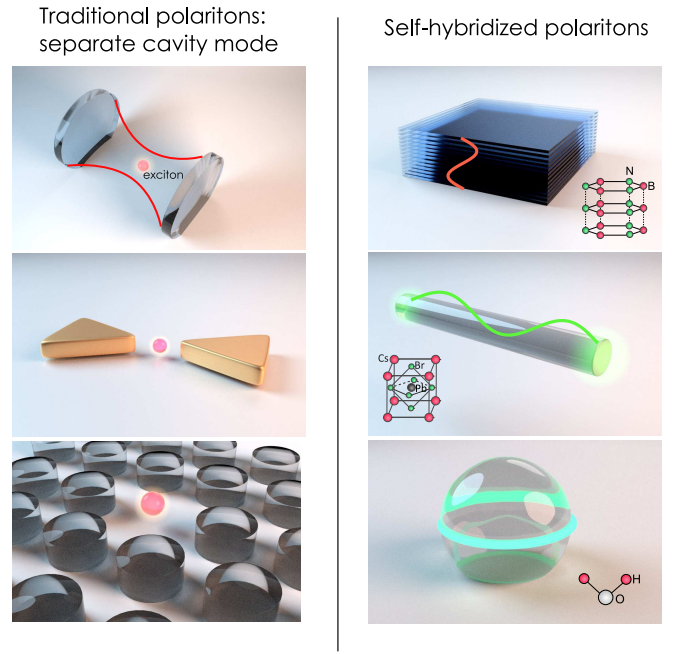}
    \caption{On the left: examples of architectures for traditional polaritons based on a separate cavity mode provided by a Fabry-Pérot resonator, a plasmonic resonator, or a photonic crystal cavity. On the right: examples of self-hybridized polaritons based on slabs, long nanorods, and spheres made of resonant (excitonic/Lorentzian) materials.}
\label{fig1:usualSC}
\end{figure}

Traditionally, one requires an external cavity to reach strong coupling, such as a FP cavity, a photonic crystal cavity, or a plasmonic nanocavity, schematically visualized in Fig. \ref{fig1:usualSC}(a).  Assuming that the role of the cavity is solely to confine modes, we argue that using external cavities may limit the scope of potential applications, as such cavities are impractical.\footnotetext[1]{An alternative assumption is that a two-metallic-mirrors cavity significantly modifies the density of electromagnetic states inside the cavity, as compared to the free space, by both increasing the density at resonance(s) and decreasing it away from resonance(s) over a very broad spectral range, in a spirit similar to a Faraday cage, which may affect chemistry and other material-related properties in both resonant and non-resonant ways (making this scenario different from the self-coupled polaritons discussed below). Such an assumption may be realistic and we cannot exclude it at this point. However, in what follows, we focus on the analysis of polaritonic eigenmodes of the system in the assumption that these polaritonic modes are somehow important for chemistry, as was suggested in the recent literature.\cite{thomas2016ground} }\footnotemark[1] There also exist a number of so-called ``open'' cavities, which might be more accessible and useful,\cite{Munkhbat2018, torma2014strong, Baranov2018} but also require an external object to provide the optical mode confinement. Here, we show that to achieve strong coupling, the cavity does not need to be external. Instead, electronic or vibrational excitations in bulk or nanostructured materials can \emph{self-couple} to an optical mode sustained by its own geometry, as illustrated in Fig. 1(b). We consider a generic medium with a Lorentz dielectric function, motivated by the fact that optical properties of most materials can be modelled as such.

It is well known that objects with refractive index different from their surrounding can sustain localized optical eigenmodes.\cite{jackson_electodyn} The characteristics of the modes depend on the geometry and refractive indices of the involved objects and its surroundings. Once such an optical resonance is found at a proximity of the material resonance, these optical and material resonances can hybridize and, provided the oscillator strength of the transition is high, give rise to polaritons. The Rabi splitting in such a self-hybridized polariton can approach (but cannot exceed) the so-called bulk polariton splitting, which is a function of oscillator strength only and does not depend on parameters of the cavity. Such self-coupled or cavity-free polaritons have been realized in many relevant systems, such as slabs of hexagonal boron nitride (hBN), transition metal dichalcogenides, perovskites, and others.\cite{Platts2009, chen2016light, Zhang2018, tiguntseva2018tunable, Munkhbat2019, Verre2019, chen2019resonant} However, their relevance for the polaritonic chemistry and modified material properties has not been raised in the community. 

Originally, the problem of material resonance from the quantum optical light-matter coupling perspective has been solved by Hopfield in his seminal work in 1958.\cite{hopfield1958theory} Hopfield considered a three-dimensional (3D) continuous semiconductor and modelled it using a quantum coupled harmonic oscillator model, but did not consider materials shaped into a specific object(s) that can sustain a set of discrete well-defined electromagnetic modes.

As we show here, cavity-free polaritons are found in a great variety of structures, ranging from bulk materials to spherical micro droplets, Fig. 1(b). The sole existence and apparent abundance of such cavity-free polaritons questions their applicability in polaritonic chemistry and, more generally, in any polariton-induced modifications of material-related properties.\cite{peters2019control}

%%%%%%%%%%%%%%%%%%%%%%%%%%%%
% Standard Polaritons%
%%%%%%%%%%%%%%%%%%%%%%%%%%%%

\section{Revisiting standard cavity-emitter polaritons}

We start by revisiting the standard light-matter interaction Hamiltonian in Hopfield's formulation.\cite{hopfield1958theory} This formulation is essentially a coupled harmonic oscillator problem approached from the quantum optical perspective. We note that this is not exactly the same model as the famous Jaynes-Cummings or quantum Rabi models, which deal with quantum emitters (QEs) coupled to a cavity mode. However, within the weak excitation regime, this formulation is sufficiently appropriate even for QEs, and even more within the scope of this study since we focus on macroscopic Lorentz-resonant descriptions. 

The full Hopfield Hamiltonian for a cavity mode oscillating at a frequency $\omega_c$ interacting at a rate $g$ with a material transition at a frequency $\omega_0$, written in the Coulomb gauge and including the diamagnetic term, reads:

\begin{multline} \label{Eq:Hopfield}
    \hat{H}= \hbar \omega_0 \hat{b}^{\dagger} \hat{b} + \hbar \omega_c \hat{a}^{\dagger} \hat{a} +  \\
    \hbar g (\hat{a}^{\dagger} + \hat{a})(\hat{b}^{\dagger}+\hat{b}) + 
    \frac{\hbar g^2}{\omega_{0}}(\hat{a}^{\dagger} + \hat{a})^2,
\end{multline}

where $\hat{b}$ and $\hat{a}$ are the annihilation operators of the material transition and the cavity mode, respectively, and $g$ is the coupling constant.\cite{Baranov2020} Two eigenvalues (resonant transitions) of this Hamiltonian are given by:

\begin{widetext}
\begin{equation}\label{Eq:polaritons}
    \omega_{\pm} = \frac{ \sqrt{\omega_c^2 + 4g^2 \omega_c / \omega_0 + \omega_0^2 \pm \sqrt{(\omega_c^2 + 4g^2 \omega_c/\omega_0 +  \omega_0^2)^2 - 4\omega_c^2 \omega_0^2 }}}{\sqrt{2}},
\end{equation}
\end{widetext}

With the material oscillators homogeneously occupying the cavity mode volume $V$, the coupling strength $ g=\mu \sqrt{\rho V/3} \mathcal{E}_{vac} \omega_0 / \omega_c$ (with the factor $1/3$ accounting for their isotropic orientation). Thus, the coupling strength depends on the transition dipole moment of the material transition, $\mu$, and the volume density of these dipoles (oscillators), $\rho$ as well as the vacuum electric field of the cavity, $\mathcal{E}_{vac}=\sqrt{ \hbar \omega_c / 2 \varepsilon_\infty \varepsilon_0 V}$. When the coupling strength is large enough, two polaritons with different energies (Eq. \ref{Eq:polaritons}) are formed, which has been explained in great detail before.\cite{Baranov2018, khitrova2006vacuum} 

From the above description, one can deduce that two main ingredients are necessary for achieving strong coupling: an optical cavity with a large quality factor ($Q$) and a small mode volume ($V$) and an electronic (or vibrational) transition with a high transition dipole moment.
Following such suggestions, there has been extensive research to obtain the best optical cavities, ranging from high-finesse FP cavities composed of distributed Bragg reflector (DBR) mirrors to plasmonic nanocavities. Many of these cavities are covered by Baranov et al. \cite{Baranov2018} and some are depicted in Figure \ref{fig1:usualSC}.

In the examples above, we see that a stand-alone cavity is important for realizing the standard cavity-quantum electrodynamics (QED) scenario, that is, the case of a single-QE strongly coupled to an external optical cavity. Such a scenario is challenging to realize with the cavity-free approach, described below. Moreover, a single QE problem can not be treated in terms of a macroscopic Lorentz dielectric functions even with the Hopfield Hamiltonian, as we do below, but instead requires a true cavity-QED approach, which we do not focus on here (these approaches are extensively covered in the literature \cite{torma2014strong, FriskKockum2019}).

%%%%%%%%%%%%%%%%%%%%%%%%%%%%%
% Self-hybridized %
%%%%%%%%%%%%%%%%%%%%%%%%%%%%%

\section{Self-hybridized polaritons}

An optical mode is a concept that is not limited to the optical resonators and cavities discussed above. An optical mode is a solution of the source-less Maxwell equations in a given geometry. Thus, a cavity can be made of any material, including the material that makes up the relevant electronic or vibrational transitions themselves. These are precisely the scenarios we focus on in the following sections.

%%%%%%%%%%%%%%%%%%%%%%%%%%%%%
% Bulk Polaritons %
%%%%%%%%%%%%%%%%%%%%%%%%%%%%%

\subsection{3D case: bulk polaritons}
Let us start by revisiting bulk polaritons, hybrid light-matter states of an unbounded dielectric medium homogeneously filled with resonant transitions. This situation has been described by Hopfield in the celebrated 1958 paper \cite{hopfield1958theory} (see also Mills et al. \cite{Mills1974}). Instead of a localized eigenmode of a cavity, the photonic modes in this scenario are represented by plane waves propagating in the unbounded dielectric medium with permittivity $\varepsilon_{\infty}$. Those waves have a continuous spectrum $\omega = kc/n$ with $k$ being the vacuum wave vector, $c$ the speed of light, and $n=\sqrt{\varepsilon_{\infty}}$ the refractive index of the medium.

In the classical electromagnetic formalism, an isotropic resonant medium can be described by a Lorentzian permittivity:
\begin{equation} \label{Eq:Lor}
    \varepsilon(\omega) = \varepsilon_{\infty}+f \frac{\omega_P^2}{\omega_0^2-\omega^2 - i\gamma \omega}
\end{equation}
with $\omega_P=\sqrt{\rho e^2/3\varepsilon_0 m}$ being the plasma frequency, where $\rho$ is the volume density of oscillators and $1/3$ accounts for their isotropic orientation, $e$ and $m$ are, respectively, the electron charge and mass and $\omega_0$ and $\gamma$ are the resonance frequency and linewidth, and $f$ the oscillator strength. The latter is expressed via the microscopic parameters of the medium as $f = 2  \frac{ m \omega_0}{e^2 \hbar}  \left| \mu \right|^2$, where $\mu$ is the transition dipole moment of the material oscillator (see Methods).
For simplicity, in Eq. (\ref{Eq:Lor}) and throughout this study (except for the case of hBN), we deal with only one resonance with an isotropic random orientation of transitions. More complicated cases, including multiple Lorentz resonances and anisotropy, naturally follow from this simplified model.  

Bulk polaritons correspond to roots of the dispersion equation:
\begin{equation}  \label{Eq:DispBulk}
    kc-\omega \sqrt{\varepsilon(\omega)}=0.
\end{equation}
One can look for solutions of this equation either with complex $\omega$ and real $k$, or complex $k$ and real $\omega$. These two types are both appropriate solutions of Maxwell's equations, but suited for different purposes.\cite{Rahmeier2020} We are interested in complex-$\omega$ solutions of the dispersion equation, since they better reflect the transient decay of hybrid light-matter states of the coupled system.

Quantitative characteristics of the problem, such as the vacuum Rabi splitting and the magnitude of the polaritonic gap, can be obtained from the Hamiltonian formulation of the problem that neglects the exciton dissipation.\cite{hopfield1958theory}
Following the standard procedure of quantizing the electromagnetic field in free space, let us consider a quantization box of volume $L^3$ with periodic boundary conditions.\cite{landau1977theoretical} The vacuum electric field of a photon with frequency $\omega_c$ reads
$\mathcal{E}_{vac}=\sqrt{\hbar \omega_c / 2\varepsilon_\infty \varepsilon_0 L^3}$.
Recalling the expression for the oscillator strength $f$ and plasma frequency $\omega_P$ of a Lorentz medium, we express the resulting coupling strength as 
$g_C = (\omega_P/2) \sqrt{f \omega_0/ \varepsilon_\infty \omega_c}$.
The vacuum Rabi splitting (with zero losses) is then obtained as the difference between the two polariton energies (Eq. \ref{Eq:polaritons}) for the zero-detuned photon ($\omega_c=\omega_0$): 
\begin{multline}
    \Omega_R=\sqrt{2g^2+\omega_0^2 + 2g\sqrt{g^2 + \omega_0^2}} - \\
    \sqrt{2g^2+\omega_0^2 - 2g\sqrt{g^2 + \omega_0^2}}=2g_0,
\end{multline}
where $g_0= (\omega_P/2)\sqrt{f /\varepsilon_\infty}$ is the zero-detuning coupling strength.

\begin{figure*}[t] %%%%%%%%%%FIG.2%%%%%%%%
  \includegraphics[width=\linewidth]{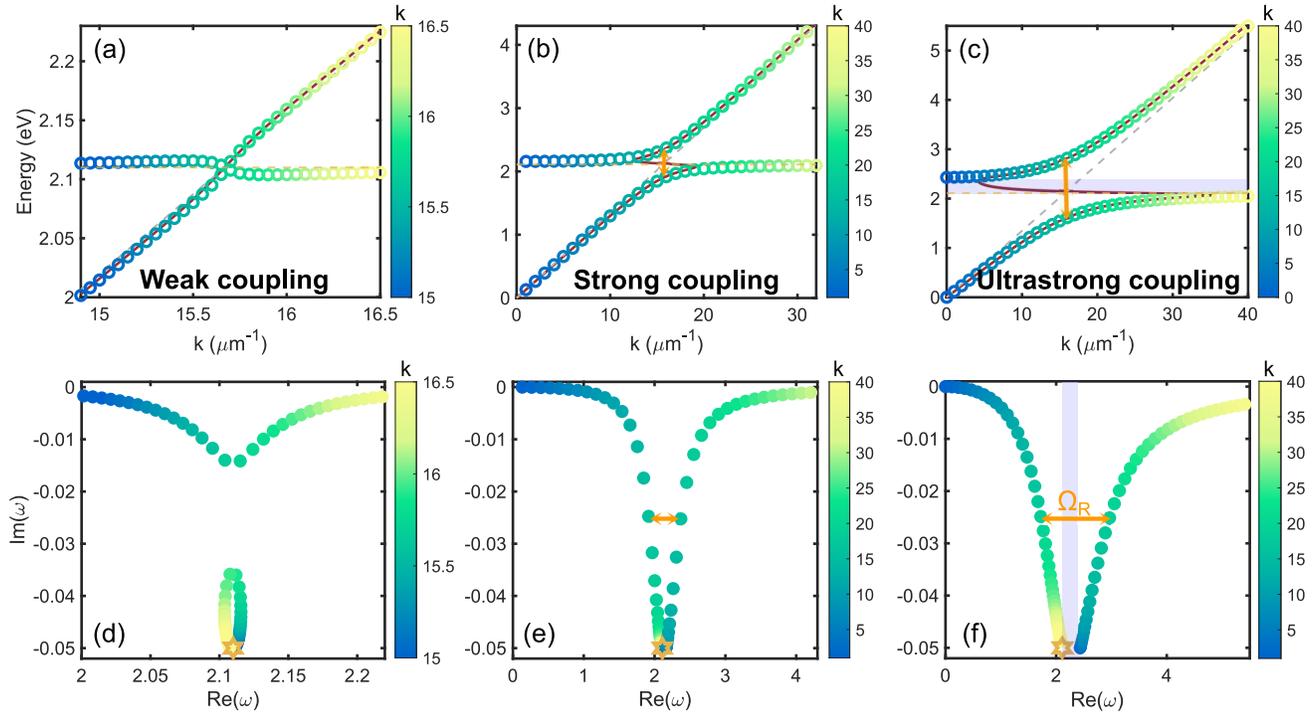}
  \caption{\textbf{Bulk polaritons}. The upper panels show the dispersion of bulk J-aggregate polaritons considering the complex-$k$ and real-$\omega$ solutions in a red solid line and the complex-$\omega$ and real-$k$ eigenmodes in circles for (a) weak coupling with $f \omega_P^2 =0.0045 ~\text{eV}^2$ (b) strong coupling with $f \omega_P^2 =0.445 ~\text{eV}^2$ and (c) ultrastrong coupling with $f \omega_P^2=3.116 ~\text{eV}^2$, which displays a polaritonic gap shaded in blue. The dashed lines denote the dispersion of the uncoupled exciton (yellow at 2.11 eV) and the light dispersion in vacuum (grey). The lower panels show the trajectories of the complex frequency roots for the same values in the 3 regimes: (d) weak (e) strong and (f) ultrastrong coupling. The location of the uncoupled exciton  ($\omega_0-i\gamma/2$) is marked as a yellow star. The Rabi splitting is marked by orange arrows. The color bar represents the value of $k$ in $\mu m^{-1}$.}
  \label{fig:2Bulk}
\end{figure*}

The upper edge of the polariton gap is obtained by calculating the upper polariton energy in the limit $k=0$ and is $\Tilde{\omega_0}=\sqrt{\omega_0^2 + 4g_0^2}$, where $\Tilde{\omega_0}$ is the re-normalized resonance frequency. The lower edge of the gap is found by calculating the lower polariton energy in the limit $k \to \infty$, and is exactly the uncoupled exciton energy $\omega_0$. This results in the polaritonic gap
\begin{equation}
    \Delta_{pol}=\sqrt{\omega_0^2 + 4g_0^2} - \omega_0,
\end{equation}
which simplifies to $\Delta_{pol} \approx 2g_0^2/\omega_0$ under the assumption of $4g_0^2/\omega_0^2\ll 1$, that holds in the standard strong coupling regime.

The presence of losses in the dielectric formalism makes the eigenfrequencies complex and changes their dispersion, however, the original Hopfield model of bulk polaritons does not include any dissipation channels. Subsequent works extended Hopfield's model to account for exciton decay by considering coupling between them and a continuum bath of harmonic oscillators.\cite{Huttner1992} Such a model predicts that bulk polaritons appear if and only if the condition
\begin{equation}
    f \omega_P^2>\varepsilon_\infty\gamma^2/4
\end{equation}
holds. It is easy to see that this condition is equivalent to 
$(\omega_P/2)\sqrt{f /\varepsilon_\infty}=g_0 > \gamma/4$, which is the strong coupling criterion in the standard Jaynes-Cummings model with a lossless cavity mode. \cite{Huttner1992,DeLiberato2017} Thus, by knowing the parameters of the material resonance in Eq.~(3), it is possible to estimate the bulk coupling strength at zero detuning $g_0$ and compare it to $\gamma/4$ in order to conclude whether or not this specific material supports bulk polaritons.

Figure \ref{fig:2Bulk}(a-c) shows the dispersion of bulk polaritons ($k$ vs real part of $\omega$) obtained for a medium described by the permittivity in Eq.~(\ref{Eq:Lor}), with typical values for TDBC J-aggregates ($\omega_0=2.11$ eV, $\gamma=0.1$ eV, $\varepsilon_{\infty}=2.15$)\cite{balasubrahmaniyam2020coupling}, and varying oscillator strengths. First, for $f \omega_P^2=0.0045 ~\text{eV}^2$, corresponding to a strongly diluted J-aggregate, we observe the mode attraction typical for the weak coupling  regime,  Fig.~\ref{fig:2Bulk}(a). In this regime the energies of the eigenmodes of the medium slightly deviate from the bare values, but their dispersions cross near the zero detuning region.

For a larger oscillator strength $f \omega_P^2=0.445 ~\text{eV}^2$, see e.g. Balasubrahmaniyam et al., \cite{balasubrahmaniyam2020coupling} one can clearly see an anticrossing between the two polaritonic modes, with the vacuum Rabi splitting around 455 meV.
We calculate the vacuum Rabi splitting as the energy separation along the real frequency axis between polaritonic modes obtained at a zero detuning between the electronic resonance of the material and the bare cavity mode (in turn, obtained by setting $f=0$). This choice corresponds to nearly $50\% / 50\%$ excitonic/cavity polariton wave-function amplitudes, and we use this definition throughout the text.
On the same plot we also show dispersion of complex-$k$/real-$\omega$ solutions (in a solid red line), found simply as $k=\sqrt{\varepsilon(\omega)}\omega/c$. The two solutions coincide away from the resonance, but differ in the vicinity of $\omega_0$. While the two branches of complex-$\omega$ solutions are disconnected, the two branches of the complex-$k$ solutions are connected by the region of anomalous dispersion.

For even larger oscillator strength, $f \omega_P^2=3.116 ~\text{eV}^2$ (we note here that this value is unrealistically large for J-aggregates and we give it here only for the sake of theoretical argument; there may be other material systems, where such oscillator strengths may be reached), the dispersion shows a larger vacuum Rabi splitting of about 1.2 eV and also displays a polaritonic gap -- a region of frequencies with no propagating modes within it, shown in Fig.~\ref{fig:2Bulk}(c). \cite{hopfield1958theory,FriskKockum2019,forn2019ultrastrong} The corresponding complex-$k$ solutions, at the same time, span over the whole range of frequencies, but become highly damped within the polariton gap by acquiring a large imaginary part of the wave vector $k$ (see Fig. S1). 

In addition to the anticrossing in the $k-\omega$ plane, it is instructive to inspect trajectories of the roots of the dispersion in Eq.~(\ref{Eq:DispBulk}) in the complex-$\omega$ plane. Fig.~\ref{fig:2Bulk}(d-f) show trajectories of the complex frequency roots for the three cases of weak, strong, and ultrastrong coupling. In the weak coupling, $f \omega_P^2 < \varepsilon_\infty(\gamma/2)^2$, the ``photonic'' root of the equation crosses the exciton position and acquires a small imaginary component in the zero detuning region, while the ``excitonic'' root (the one originating at the complex exciton frequency, marked with a yellow star in Fig.2) moves along a finite curve around $\omega_0-i\gamma/2$. However, in the strong and ultrastrong coupling regimes the low-energy solution terminates exactly at the exciton energy, while the upper solution originates on the other side of the polaritonic gap, Fig.~\ref{fig:2Bulk}(e,f).

In closing this part we want to emphasize that bulk polaritonic states and the whole anticrossing picture exist in an optically large piece of a resonant material without any external cavity.

%%%%%%%%%%%%%%%%%%%%%%%%%%%%%%%%%%%%%%%%%%%%%%%%%%%%%%%%%%%%
\subsection{2D case: polaritons in Lorentz slabs} \label{subsec:2D}

The examples given above clearly illustrate that a large (compared to the resonant wavelength) unstructured excitonic or phononic material already possesses polaritonic modes, provided that the oscillator strength of the material exceeds the threshold value. Reducing the dimension of the system (by breaking the translational invariance in one, two, or all three dimensions) does not break this picture: the reduced system still possesses optical modes (either complex-frequency resonant states, or real-frequency guided modes) that couple with the material resonance.

\begin{figure*} %%%%%%FIG.3%%%%%%%%%%%%%%
  \includegraphics[width=\linewidth]{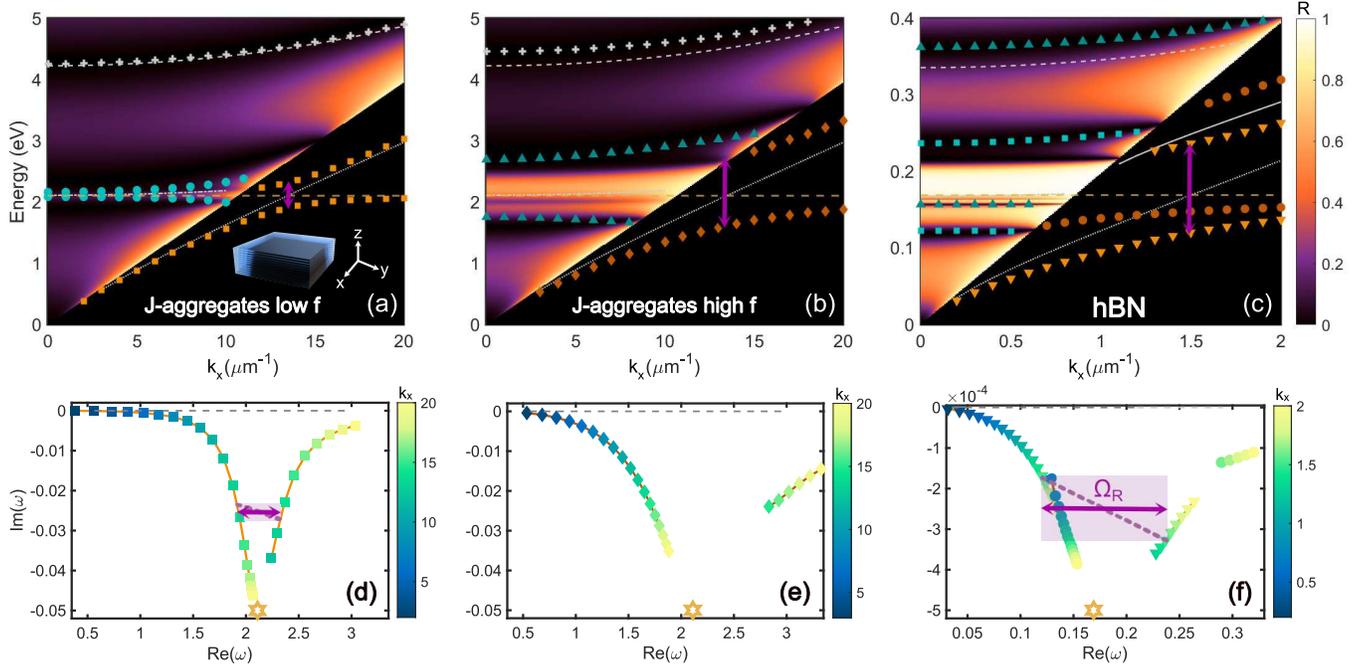}
  \caption{\textbf{Polaritons in dielectric slabs}. The upper panels show dispersion of the Fabry-Pérot (FP) and guided modes over the reflection spectra of TE polarized light of (a) a 200 nm thick J-aggregate slab with $f \omega_P^2 =0.445 ~\text{eV}^2$ (b) with $f \omega_P^2 =3.116 ~\text{eV}^2$ and (c) a 1.75 $\mu$m thick slab of hBN with $f \omega_P^2 =0.064 ~\text{eV}^2$. The lower panels show the trajectories, in the complex-$\omega$ plane, of the polaritons resulting of the 1$^\mathrm{st}$ guided mode of the slab (TE$_1$) hybridizing with the resonance in the same materials as above: (d) J-aggregates with low $f$, $\square$ (e) J-aggregates with high $f$, $\Diamond$, and (f) hBN, $\triangledown$ for hybridization with TE$_1$ and $\circ$ with TE$_2$. The colorbar represents $k_x$. The Rabi splitting with TE$_1$ is marked with a purple arrow. The uncoupled modes are shown in dashed and dotted lines.}
  \label{fig:3Slabs}
\end{figure*}

Let us first reduce the bulk material in one dimension (along the $z$-axis, to be specific) and consider a slab with a Lorentzian permittivity $\varepsilon(\omega)$ in vacuum. The eigenmodes of this system can be quantified by the (real) in-plane momentum $k_x, k_y$.\cite{Armitage2014} In the absence of the material resonance, the modes above the light-line are FP resonances with complex frequencies (not to be confused with real-frequency leaky modes);\cite{Lalanne2018} the modes below the light-line are proper waveguide modes with real frequencies.\footnotemark[2]\footnotetext[2]{Strictly speaking, radiating complex-frequency eigenmodes of the slab extend below the light-line, and only indirectly connect to the waveguide modes via a branch of the improper mode dispersion.\cite{Abdrabou2019} However, this effect is very subtle and can be ignored in the present analysis} Upon inclusion of the material resonance, it couples to modes both below and above the light-line.

Figure \ref{fig:3Slabs}(a) presents TE-polarized eigenmodes of a 200 nm thick J-aggregate slab with $f\omega_P^2 =0.445 ~\text{eV}^2$. Such J-aggregate was also analyzed in Fig. \ref{fig:2Bulk}(b,e), corresponding to the bulk strong coupling case. It shows the real part of eigenfrequencies versus the in-plane momentum $k_x$ on top of the reflection spectrum calculated above the light-line.
Eigenfrequencies of the slab are found numerically as poles of the reflection coefficient in the lower half-plane of the complex frequency (see Methods).
Remarkably, spectral positions of FP modes above the light-line do not always coincide with reflection dips, as one can see from Fig. \ref{fig:3Slabs}(a) for polaritonic modes, and Fig. S2 for bare modes (f=0) of a dielectric slab. This behavior comes from non-resonant contributions of other poles of the dielectric slab, in particular, the fundamental waveguide mode present at any $k_x$. \cite{Armitage2014} This mismatch should be always kept in mind in analysis of experimental or simulated spectra of coupled exciton-polariton systems.

The 200 nm thick slab with $f \omega_P^2 =0.445 ~\text{eV}^2$ does not show any noticeable splitting above the light-line ($\circ$ in Fig. \ref{fig:3Slabs}(a)): the coupling strength is not sufficient to overcome the large radiative decay of the FP mode. Below the light-line, however, there is a clear anti-crossing between the exciton and the TE$_1$ waveguide mode with a magnitude of Rabi splitting around 420 meV, which is smaller than the 455 meV of bulk Rabi splitting previously mentioned. 

TE-polarized eigenmodes of a 200 nm thick J-aggregate slab with a higher oscillator strength $f \omega_P^2 =3.16 ~\text{eV}^2$ demonstrates anticrossing both below and above the light-line, Fig. 3(b).  This J-aggregate was analyzed in Fig. 2(c,f), and corresponds to the bulk ultrastrong coupling case. The location of the Rabi splitting with TE$_1$ is marked with a purple arrow, but its magnitude cannot be obtained because the upper polariton has a cut-off given by the light-line and the corresponding guided eigenmode does not exist.

Next, we demonstrate self-hybridized polaritons in another practically relevant system: a slab of hexagonal boron nitride (hBN).  hBN exhibits two prominent vibrational modes at 169 and 95 meV, respectively.\cite{Caldwell2015} We will focus on TE-polarized modes, which couple only with the in-plane vibrations of hBN at 169 meV. The in-plane permittivity component of hBN can be described by a single Lorentz term with $\varepsilon_{\infty}=4.45$, $f\omega_P^2 = 0.064 ~\text{eV}^2$ , and $\gamma=1$ meV, which approximates the experimental data in the relevant spectral range (Fig. S1(g)).\cite{Caldwell2015}

Fig.~\ref{fig:3Slabs}(c) shows eigenfrequencies of TE-polarized eigenmodes of a 1.75 $\mu$m thick hBN slab. Clearly, there are anticrossings both below ($\circ$ and $\triangledown$) and above ($\square$ and $\vartriangle$) the light-line thanks to the large oscillator strength of hBN's vibrational transition. For the same reason, we also observe the lower polariton branch of the 2$^\mathrm{nd}$ FP mode above the light-line, shown in blue $\vartriangle$, and lower polariton of the TE$_2$ waveguide mode below the light-line (orange $\circ$). The Rabi splitting of the TE$_1$ waveguide mode, which occurs at about $k_x=1.5$ $\mu m^{-1}$, reaches a value of $\approx 116$ meV, which is also below the bulk Rabi splitting, $2g_0 \approx 120$ meV. Remarkably, below the light-line, the two fundamental modes -- TE$_1$ and TE$_2$ -- show pronounced anticrossing, even though the TE$_2$ mode dispersion in a dielectric slab without the exciton does not approach the exciton energy (see Fig. 3(c)).

This behavior becomes more evident when we look at trajectories of the eigenmodes' poles of the slab in the complex-$\omega$ plane. It turns out, that even for a vanishingly small oscillator strength $f \omega_P^2$ there are countably many poles in an arbitrarily small vicinity of the complex exciton frequency $\omega_0-i\gamma/2$ (see Fig. S3(b)), where the Lorentz permittivity $\varepsilon(\omega)$ has a simple pole (shown in Fig. S3(c)).\cite{Broer2009, Muljarov2016} Each of these poles represents an eigenmode of the Lorentzian slab: either a radiative mode, or a localized waveguide mode, depending on the position of the pole frequency with respect to the complex plane cut.

The origin of this cluster of poles can be appreciated by considering the Lorentzian permittivity $\varepsilon$ at complex frequencies, Fig. S3(c). One can clearly see that just below the complex exciton frequency, $\omega_0-i\gamma/2$, the permittivity takes arbitrarily large values, thus, in theory, allowing for the existence of higher-order FP and waveguide modes, that are otherwise inaccessible at real frequencies, where the permittivity is bounded by $|\varepsilon(\omega)| \leq \varepsilon_{\infty}+f\omega_P^2/\gamma \omega_0$.

Because these poles are located so close, they usually merge into a single spectral feature in real-frequency spectral responses, often attributed to ``uncoupled molecules''.\cite{antosiewicz2014plasmon, fauche2017plasmonic} In fact, as the complex frequency plane reveals, this feature stems from a whole new multitude of eigenmodes located in the small vicinity of the complex exciton energy (Fig. S3). Of course, only a finite number of these eigenmodes are physical, since the permittivity becomes highly non-local and loses its physical meaning as soon as the wavelength inside the material $\lambda/ \operatorname{Re}(\sqrt{\varepsilon(\omega)})$ at complex $\omega$ approaches the microscopic length scale of the medium near the pole.

\begin{figure*}[t] %%%% FIG.4 Cylinders %%%%%
  \includegraphics[width=\linewidth]{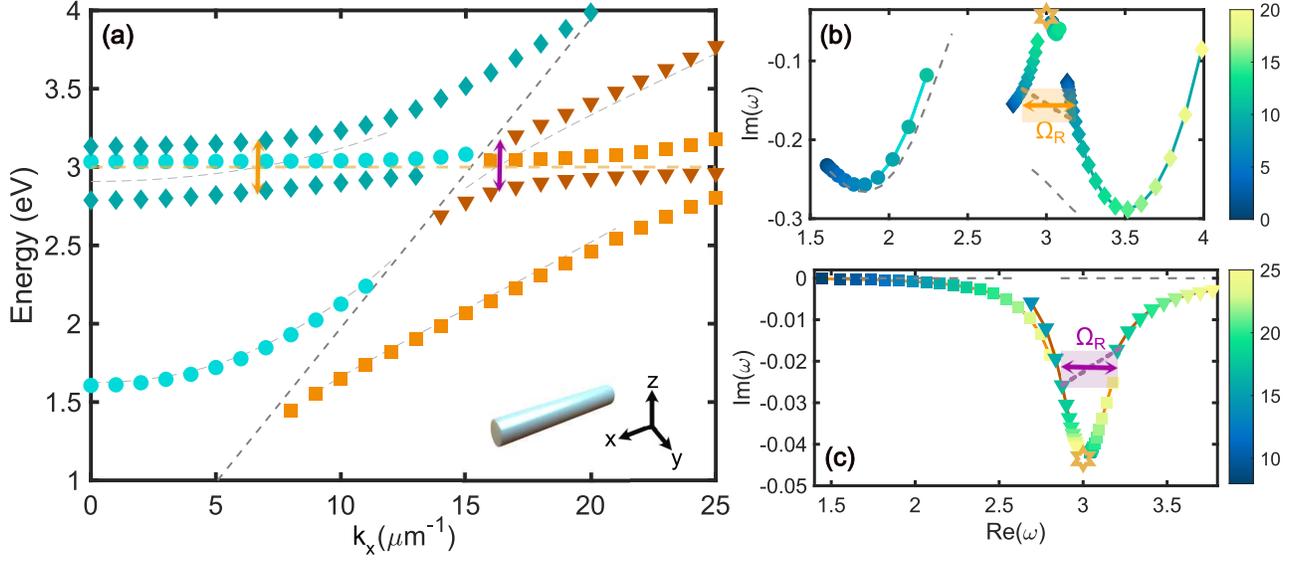}
  \caption{\emph{\textbf{Polaritons in infinite circular cylinders}}. (a) Dispersion of the polaritons given wtih the TM radiating modes of  CsPbCl$_3$ cylinder, with a radius of 250 nm, above (1$^\mathrm{st}$ mode in $\circ$, 2$^\mathrm{nd}$ mode in $\Diamond$) and waveguide modes below the light-line (1$^\mathrm{st}$ mode in $\square$ and 2$^\mathrm{nd}$ in $\triangledown$). Trajectories of the hybrid modes with (b) radiating and (c) waveguide modes in the complex-$\omega$ plane. The colorbar shows the value of $k_x$ and the colored arrows show the Rabi splitting above (orange) and below (purple) the light-line. The uncoupled optical modes are shown in dashed lines.}
  \label{fig:4cylinder}
\end{figure*}

Complex-plane trajectories also clearly reveal the Rabi splitting. Above the light-line trajectories of the eigenmodes are limited by the low frequency cut-off on the one side, and by the light-line on the other side, and for this reason are difficult to interpret (Fig. S4). Below the light-line, however, only the upper polariton branches are bounded by the light-line on the low energy side, Fig. \ref{fig:3Slabs}(d), and the resulting trajectories are very similar to those for bulk polaritons (see Fig. 2(e)).
Qualitatively similar pictures of anticrossing in the $\omega-k_x$ space and in the complex-$\omega$ plane are found for TM-polarized eigenmodes (Fig. S5).
%%%%%%%%%

%%%%%%%%%%%%%%%

%%%Electrically thin
Since the waveguide modes of a dielectric slab have virtually infinite lifetime in contrast to radiative FP modes, anticrossing below the light-line appears at a much smaller oscillator strength. For an electrically thin film we can establish an approximate analytical criterion of strong coupling for waveguide modes. 
Assuming an electrically thin dielectric film without the Lorentzian transition ($k_0L\sqrt{\varepsilon_\infty}\ll1$, where $L$ is the thickness of the film), the dispersion of the TE$_1$ waveguide is approximately 
\begin{equation}
    k_x \approx k_0 (1+\delta k),~~ \delta k= (k_0 L)^2 (\varepsilon_{\infty} - 1)/8,
\end{equation}
with $\delta k$ being a small parameter for a thin film (see Methods).
The vacuum field of the bare TE waveguide mode in the center of the dielectric slab can be estimated as $\mathcal{E}_{vac}=\sqrt{\hbar\omega_c/2\varepsilon_0 a^2 W}$, where $W=1/\operatorname{Im}(k_{z,1}) \approx \frac{1}{k_0 \sqrt{2 \delta k}}$ is the waveguide mode's effective width. Multiplying the vacuum field by the square root of the number of oscillators $\rho a^2 L$ in the volume $a^2 L$ and the dipole moment $\mu$, we find the coupling strength (at zero detuning, $\omega_{c}=\omega_0, k_0=\omega_0/c$):
\begin{equation}
    g=\frac{\omega_P}{2}(k_0L) \sqrt{ \frac{f \sqrt{\varepsilon_\infty- 1 }}{2}}.
\end{equation}
Comparing the above $g$ to $\gamma/4$, we find the threshold oscillator strength required for anticrossing below the light-line in a dielectric film:
\begin{equation}
    f\omega_P^2 > \frac{ \gamma^2}{2(k_0 L)^2 \sqrt{\varepsilon_\infty - 1}}.
\end{equation}
This condition may help estimate the minimal thickness $L$ of an excitonic film required for the emergence of polaritonic modes below the light-line for a given oscillator strength.

%%%%%%%%%%%%%%%%%%%%%%%%%%%%%%%%%%%%%%%%%%%%%%%
\subsection{1D case: polaritons in long  circular cylinders} \label{subsec:1D}

By reducing the system in one more dimension (say, along the $y$-axis) we arrive at one possessing translational invariance only in a single direction (along the $x$-axis). Thanks to this symmetry, eigenmodes of the one-dimensional system are characterized by a real-valued momentum $k_x$. Similarly to the 2D case, eigenmodes with $k_x>\omega/c$ are guided modes with real frequencies; eigenmodes with $k_x<\omega/c$ are radiating resonances with complex frequencies. 

For simplicity, let us focus on 1D systems with a circular cross-section. The states with $k_x=0$ (normal-incidence resonances of the cylinder) are classified by their polarization state (TE or TM) and the azimuthal number $m$. \cite{Doost2013} 
Eigenmodes with $k_x \ne 0$ split into TE and TM-polarized solutions only for $m=0$ (the monopole harmonic).\cite{ishimaru2017electromagnetic} For $m>0$, TE and TM polarizations couple, giving rise to hybrid modes notated EHnm and HEnm. The HE11 mode (the dipole mode) is the only mode of a circular cylindrical dielectric waveguide not having a cut-off\cite{ishimaru2017electromagnetic} (see $\triangledown$ in Fig. S6(a,c)).

\begin{figure*}[t] %%%%%%FIG.5%%%%%%
\includegraphics[width=\linewidth]{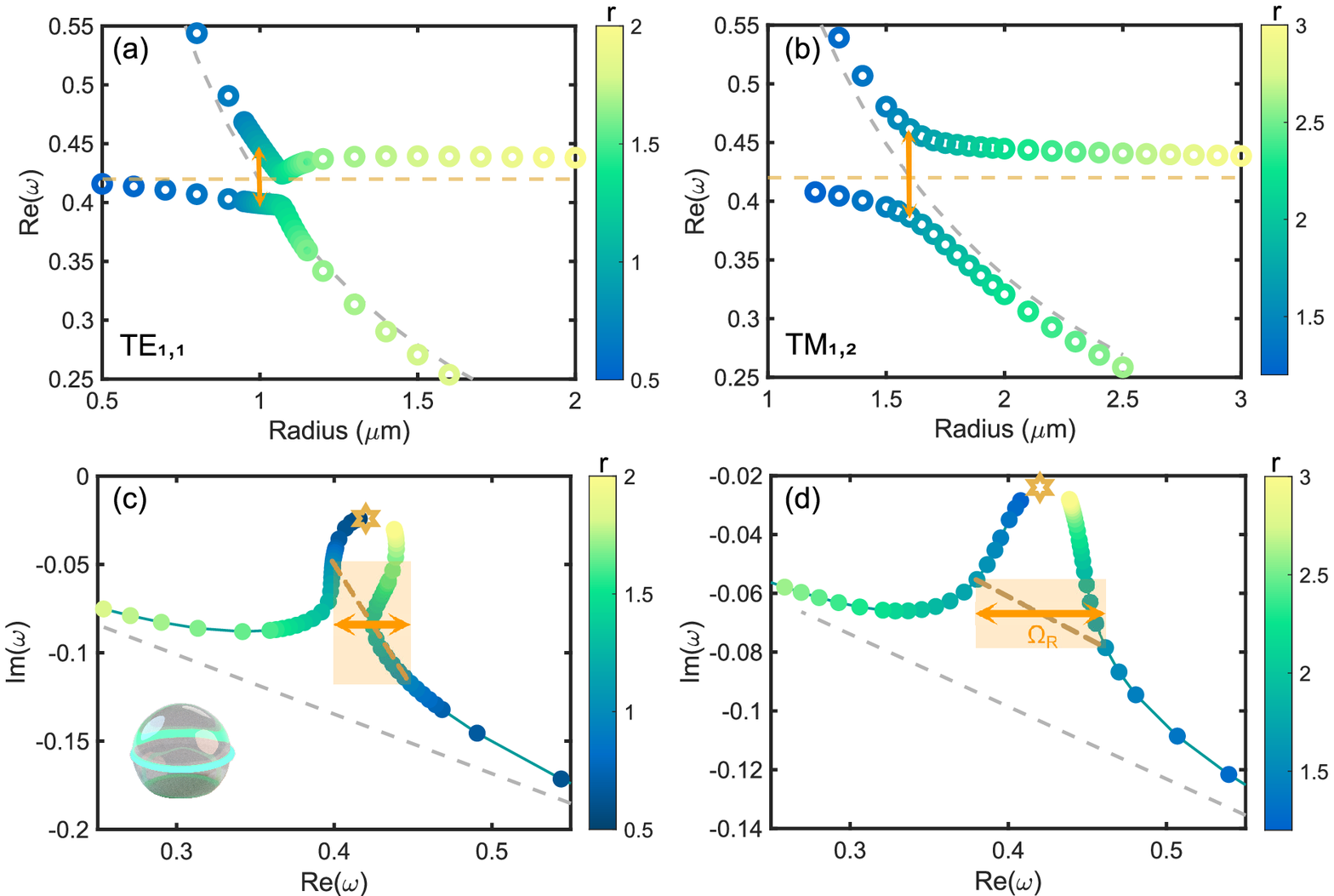}
  \caption{\emph{\textbf{Polaritons in water spheres}}. The upper panels show eigenfrequencies dependence on the radius of a water sphere for (a) TE$_{1,1}$ and (b) TM$_{1,2}$ modes. The lower panels show the trajectories of the complex$-\omega$ roots for the same modes (c) TE$_{1,1}$ and (d) TM$_{1,2}$ by varying the radius (shown in the colormap). The poles at zero detuning are noted in the trajectories as the vertices of the orange rectangle joined by the dashed line. The Rabi splitting is the width of such rectangle in $\operatorname{Re}(\omega)$.}
  \label{fig:5Spheres}
\end{figure*}

As a practical example, we demonstrate self-hybridized polaritons in infinite circular cylinders of a perovskite CsPbCl$_3$. This material is described by a Lorentz permittivity with $\varepsilon_\infty=3.7$, $\omega_0=3$ eV, $\gamma=87$ meV, and  $f \omega_P^2 =0.54~\text{eV}^2$ that approximates the experimental data \cite{tiguntseva2018tunable} (see Fig. S1(a)) . Polaritons and polaritonic emission in long whiskers made of perovskites of various families have been demonstrated in a number of works.\cite{Zhang2018, evans2018continuous}

Fig.~4 presents frequencies of TM-polarized monopole ($m=0$) eigenmodes for CsPbCl$_3$ infinitely long circular cylinder with a radius of 250 nm. The resulting spectrum clearly shows anticrossing above the light-line (involving the 2$^\mathrm{nd}$ order radiating mode) and below the light-line (involving the TM$_{0,1}$ waveguide mode). The anticrossing below the light-line features a Rabi splitting of about 312 meV, which compares to bulk Rabi splitting of CsPbCl$_3$ of 382 meV. Complex-$\omega$ trajectories of eigenfrequencies exhibit a familiar behavior with upward frequency pulling to the real axis above the light-line, and downward pulling below the light-line, Fig.~4(b,c). Qualitatively similar eigenfrequencies are obtained for TE-polarized monopole modes (Fig. S6), and for the dipole TEM modes (Fig. S7).
 
%%%%%%%%%%%%%%%%%%%%%%%%%%%%%%%%%%%%%%%%%%%%%%%%%%%%%%%%%%%%%%%%%

\subsection{0D case: polaritons in spheres} \label{subsec:0D}

Finally, if we reduce the system in one more dimension, we end up with a compact scatterer supporting a discreet spectrum of localized eigenmodes.\cite{conwell1984resonant, doost2014resonant} Because of the lack of translational invariance in any direction, all eigenmodes of the scatterer are complex-frequency radiating solutions. \footnotetext[3]{Except for the rare case of embedded eigenstates, which can occur in 0D systems only if it contains $\varepsilon=0$ materials \cite{Silveirinha2014}}\footnotemark[3]

To keep the analysis simple, we consider the most symmetric compact scatterer: a sphere, whose eigenmodes are the well-known Mie resonances.\cite{Bohren2004} These resonances split into TE- and TM-polarized modes, which are further classified by the angular index $l$, and the radial number $N$. $l=1$ modes are usually referred to as the dipole ones, $l=2$ as the quadrupole ones, and so on.
A detailed description of whispering gallery exciton-polaritons in GaAs spheres has been presented by Platts et al.\cite{Platts2009}
Because of the lack of translational invariance, the anticrossing in spheres cannot be seen by scanning the eigenmode's momentum $k$. Instead, one usually varies the radius of the sphere, what causes the eigenmodes to move in the $\omega$-space.

As a practically interesting example of self-hybridized polaritons in spheres, we consider liquid water droplets. Liquid water has a strong vibrational resonance around 420 meV with the oscillator strength of $f \omega_P^2 =0.035 ~\text{eV}^2$.\cite{segelstein1981complex} Therefore, one may naturally expect formation of vibrational polaritons in Mie resonant micron-sized water droplets.
We describe water by a Lorentz permittivity with $\varepsilon_\infty=1.75$, $\omega_0=0.42$ eV, $\gamma=0.048$ eV, and  $f \omega_P^2 =0.035 ~\text{eV}^2$ that approximates the experimental data, see Fig. S1(d).\cite{segelstein1981complex}

Fig. \ref{fig:5Spheres}(a) shows eigenfrequencies of TE$_{1,1}$ (magnetic dipole) eigenmodes of a water sphere as function of its radius $r$. 
The trajectory of the TE$_{1,1}$ eigenmode (the first index indicates the angular number of the mode $l$, and the second index stands for the radial number of the mode $N$) shows a clear anticrossing with a magnitude of about 30~meV for a sphere with $r\approx1.1$~$\mu$m (Fig. \ref{fig:5Spheres}(c)).
This value is significantly lower than the corresponding bulk Rabi splitting for water $\omega_P\sqrt{f /\varepsilon_\infty} \approx 100$ meV due to high radiative loss of the TE$_{1,1}$ mode. The small value of the Rabi splitting in this case is also manifested in the shape of the dispersion curves, similar to the exceptional point (EP) regime with the square root dependence of eigenenergies on the perturbation parameter (such as the radius).\cite{Ozdemir2019}

We find that 1.1 $\mu$m is the smallest radius for which polaritonic modes appear in water spheres. Although the TM$_{1,1}$ (electric dipole) eigenmode of smaller $r \approx 600$ nm spheres with $\varepsilon_{\infty}=1.75$ matches the exciton position, it has much lower $Q$-factor, which prevents strong coupling in spheres of this size (Fig. S8). 

Nevertheless, the TM$_{1,2}$ mode of $r \approx 1.7$~$\mu$m spheres does demonstrate a clear anticrossing with the water vibration, Fig. \ref{fig:5Spheres}(b,d), producing a Rabi splitting of about 70 meV, which approaches the bulk splitting of 100 meV.
Complex-$\omega$ trajectories in both cases reveal the characteristic eigenfrequency pulling to the real axis, Fig. 5(c,d), originating from large radiative losses of the bare optical mode compared to the exciton linewidth.
With increasing radius of the sphere new eigenmodes corresponding to higher orbital and radial numbers (and having higher $Q$-factor) will match the vibrational resonance position and show additional anticrossings and polaritonic modes.

We further investigate how the attainable Rabi splitting in water spheres depends on the eigenmode polarization and the orbital number. Fig. \ref{fig:6SummaryLimit}(a) shows the Rabi splitting for TE and TM eigenmodes as a function of the orbital number $l$. One can see that the Rabi splitting of TE modes systematically exceeds that of TM modes for all orbital numbers, but both gradually approach the bulk water splitting, $2g_0$, which for water is $\sim 100$ meV (see table S1). 

Coincidentally, micron-sized droplets of liquid water that support vibrational self-polaritons are quite ubiquitous in nature. For example, a common size of water droplets in fogs, mists, clouds, or steams falls exactly into the range of 1-10 $\mu m$ that is needed for the formation of vibrational Mie self-polaritons. Due to the surface tension these droplets adopt spherical geometry, thus forming an ideal natural platform for observation of vibrational self-polaritons in strong and even ultrastrong coupling regime. Note, that the bulk polariton splitting of liquid water falls into the vibrational ultrastrong coupling regime, since $g_0^{\mathrm{H}_{2}\mathrm{O}}\approx 0.05$ eV and $\eta_0^{\mathrm{H}_{2}\mathrm{O}}\approx 0.12>0.1$, which satisfies the ultrastrong coupling criterion.\cite{li2020cavity}

%%%%%%%%%%%%%%%%%%%%%%%%%%%%%%%%%%%%%%%%%%%%%%%%%%%%%%%%%%%%%%%%%
\section{Discussion}

\subsection{Critical size for the cavity-free polariton formation}
\begin{figure*} %%%Limit
  \includegraphics[width=\linewidth]{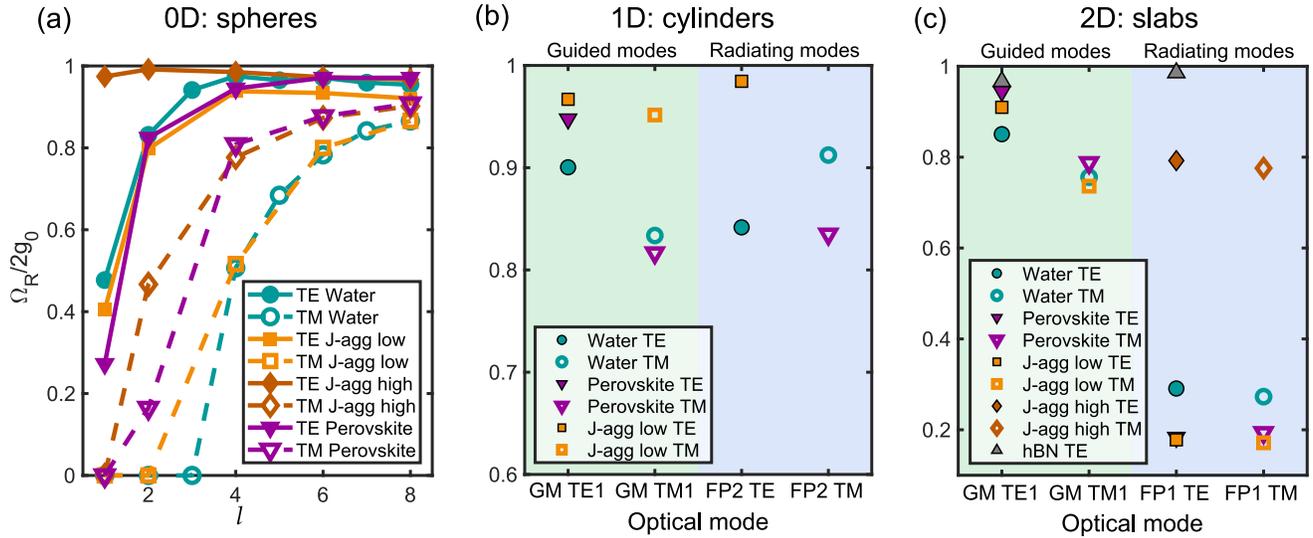}
  \caption{\textbf{Limit of Rabi splitting.} Normalized Rabi splitting for all the materials as a function of the optical mode for all the geometries discussed above: (a) spheres, (b) infinite cylinders and (c) slabs. The guided modes are shaded in green and the radiating modes are shaded in blue. The dimensions of the structures and the values of $2g_0$ are found in section V of the S.I.}
  \label{fig:6SummaryLimit}
\end{figure*}

In this section, we turn our attention to the apparent fact that nanostructures supporting cavity-free polaritons have a certain critical size, below which no polaritonic behavior is observed. For example, water spheres below $\approx 1~ \mu$m in radius will not support polaritons. Such size-limiting behavior is, of course, not universal to water and should be observed in all Lorentz materials (e.g. $r_{min} \approx 98$ nm for perovskite and $\approx 180$ nm for J-aggregates in Fig.\ref{fig:6SummaryLimit}(a)). Neither is it exclusive to the spherical geometry. For dielectric slabs, where the fundamental waveguide modes do not have a cut-off, the physical reason for the critical size is the delocalization of the electric field of the mode and a resulting weak coupling strength.
For circular cylinders (and their monopole modes) and spheres the size bound stems from the fact that the fundamental eigenmodes of electrically small systems are far blue-detuned with respect to the material transition resonance, in which case efficient hybridization is not possible. Finally, we would like to mention, that in the opposite case of exceedingly large 0D-2D objects, the result is trivial and reduces to the bulk case, which has no critical upper limit.   

From the material-science perspective the existence of the critical size suggests that the polaritonic behavior and hence modified material properties are supported only within a certain size range, given by the material resonant properties, its surrounding(s), and its shape. From the computational point of view, it is also interesting to note that the particles of about the critical size, although are on the order of several tens to several hundreds of nanometers, contain a large number of oscillators (atoms or molecules). Thus, it is challenging to self-consistently capture these collective cavity-free polaritons using atomistic methods, such as density functional theory \cite{rossi2019strong,schafer2019modification}, due to their high computational cost. 

\subsection{The limit of Rabi splitting}

The diversity of realizations of strong coupling, both using cavities (Fig. \ref{fig1:usualSC}) and in the cavity-free format (Figs. 2-5), raises a question about the limit of $g$ and $\Omega_R$ that one can observe using Lorentz materials. A particularly important question is whether it is the material or the cavity, which plays the decisive role in this limit. 

To shed light on this issue, we compare the bulk polariton coupling strength to the other realizations for the specific Lorentz material considered above, Fig. 6. From the presented data it appears that the lossless bulk polariton splitting, $2g_0 = \omega_P\sqrt{f /\varepsilon_\infty}$ (see table S1 for the values for each material), indeed establishes an ultimate limit on vacuum Rabi splitting achievable with a given Lorentz material with \emph{any} cavity mode.
Indeed, a cavity of mode volume $V$ supporting a localized eigenmode can host only as many as $N=V\rho$ transitions. Assuming they all couple to the mode with an identical coupling strength, which is certainly an overestimation, the resulting Rabi splitting is seen to be bound from top by the bulk polariton coupling. The lossless bulk polariton splitting is in turn limited from above by setting $f$=$\varepsilon_{\infty}$=1, which gives $\Omega_R^{max} = 2g_0^{max} = \omega_P$. Thus, the maximum bulk polariton splitting is principally limited from above by the volume density of oscillators -- $\rho$ (and a combination of natural constants).

It is also instructive to consider the maximum ratio between the coupling strength and the resonance frequency $\eta^{max}=g_0^{max}/\omega_0=\omega_P/2\omega_0$. This ratio plays an important role, when the coupling strength approaches the transition frequency. Specifically, when $\eta>0.1$, one usually distinguishes the ultrastrong coupling (USC) regime, while when $\eta>1$ $-$ the deep strong coupling (DSC) regime. Taking in mind the principal limits of the coupling strength, the parameter $\eta$ is seen to be intrinsically related to the ratio of plasma and resonant frequencies in Lorentz materials. This ratio can be both below one and significantly above one (in principle unbound), depending on the material. 
However, one should keep in mind that in the limit of DSC the light and matter components of the polaritonic states actually decouple, presenting nearly uncoupled optical mode and material excitation as the eigenstates of the system.\cite{DeLiberato2014} This decoupling can be seen as the result of the screening by the diamagnetic term.
In order to maintain equal cavity/matter fractions, one would have to tune the matter transition in resonance with the renormalized cavity mode.
The same decoupling occurs in the dipole gauge, where the matter resonance experiences a de-polarization shift due to dipole-dipole interaction of the electronic subsystem.\cite{Todorov2012}
In fact, considering the ratio between the coupling constant and the renormalized resonance frequency $\eta=g_0/\Tilde{\omega_0}=g_0/\sqrt{\omega_0^2 + 4g_0^2}$, one can see that the latter is principally bound by $\eta \leq 1/2$.\cite{vasanelli2016ultra} 
This hard limit indicates impossibility of reaching DSC with Lorentz media (if normalization is performed with respect to renormalized frequency).

Since the Rabi splitting is limited from top by the bulk polariton coupling strength, having a separate cavity at best allows confining polaritons in space, but does not increase the coupling strength due to the above limitation. As we showed above, the optical mode confinement can be attained without a separate cavity in the first place, with the exciton (of phonon) material itself playing the role of a confining resonator.

The bulk polariton splitting limit may serve as a practical and useful guidance for the selection of material platform in strong coupling experiments. This limit is the function of the material only, and does not depend on the nature of the cavity. For any practical realization of strong coupling the observed Rabi splitting can only approach that of the bulk, $2g_0= \omega_P \sqrt{f /\varepsilon_\infty}$, an estimation of which is straightforward (requires no complicated simulation). Thus, a material with high enough oscillator strength will naturally 
support polaritonic eigenstates in all studied 0D-3D regimes.

%%%%%%%%%%%%%%%%%%%%%%%%%%%%%%%%%%%%%%%%%%%%%%%%%%%%%%%%%%%%%%%%%%%%
%%%% CONCLUSION
%%%%%%%%%%%%%%%%%%%%%%%%%%%%%%%%%%%%%%%%%%%%%%%%%%%%%%%%%%%%%%%%%%%%

\section{Conclusion}

To conclude, we have shown that polaritonic states are natural and ubiquitous to bulk materials and nanostructures that can be described by a generic Lorentz resonance(s). Such polaritons are universally encountered in systems that do not involve any stand-alone cavities or metallic mirrors. In these cavity-free systems, boundaries of the Lorentz medium play the role of the mirrors that allow formation of well-defined optical modes, which in turn couple to the resonant transition.
We demonstrated such cavity-free polaritonic states in 3D-bulk materials, 2D-slabs, 1D-cylinders, and 0D-spheres. Concrete examples of real material systems, including spherical water droplets, perovskite nanocylinders, as well as slabs of J-aggregates and hBN were provided.

These findings might have important implications for the recently observed modification of material and chemical properties seen in electronic and vibrational strongly coupled systems. In particular, from the classical electromagnetism point of view, as well as the Hopfield Hamiltonian treatment, the cavity-free polaritons are \emph{no} different from the more standard microcavity polaritons containing dense molecular layers. Thus, all observations relevant for the cavity polaritons could be just as well applicable to the cavity-free polaritons presented here, provided it is indeed the polaritonic nature of those modes, which is responsible for chemistry and material science modifications (other reasons, such as usage of highly polarizable metallic mirrors as was done in nearly all experimental studies so far might also play a role\footnotemark[1]). However, in the cavity-free case, the exact polaritonic eigenmode picture depends on the size and shape of the resonant material and its environment. Thus, if we accept the hypothesis that polaritonic states themselves (and not the presence of metallic mirrors) are able to modify material and chemical properties, then these material and chemical properties should be dependent on the size and shape of this material, in agreement with the cavity-free polaritonic eigenmode picture. Clearly, such a cavity-free approach to modified material properties is beneficial from the practical point of view, as many of these structures exist naturally and are thus trivial to fabricate. A common concrete example considered here is the water droplets (encountered in fogs, mists, and clouds), which have a sharp cut-off below the radius of about 1 $\mu$m. Smaller water droplets will not support vibrational Mie self-polaritons and thus may experience different chemical properties in comparison to bigger droplets that support vibrational Mie self-polaritons and to bulk water that supports bulk vibrational polaritons. Similar consequences may apply to J-aggregates, hBN, etc., as well as to any Lorentz material with high enough oscillator strength of the relevant (electronic or vibrational) transition. This reasoning applies also to polariton-assisted modification of other material-related properties. We look forward to experimental tests of these predictions.

\section{\label{sec:methods} Methods}

\subsection{\label{sec:methods:Bulk} Permittivity of bulk medium}

The permittivity of a resonant medium containing electronic or vibrational transitions can be obtained by calculating the electric dipole polarizability of a single such transition, and combining it with the volume density of homogeneously distributed transitions $\rho$ in the medium. 

The dipolar polarizability of a two-level system with the transition dipole moment $\mathbf{\mu}$ in a weak external field can be written as :
\begin{equation}
    \hat{\alpha}_{\rm TLS} = \hat{\mathbf{f}} \frac{e^2/\varepsilon_0 m}{\omega_0^2-\omega^2 - i\gamma \omega},~~~ \hat{\mathbf{f}} = 2 \frac{ m \omega_0}{e^2 \hbar}\mathbf{\mu} \otimes \mathbf{\mu}
\end{equation}
where $\hat{\mathbf{f}}$ is the transition's oscillator strength, $m$ is the electron mass, $e$ is the electron charge, and $\otimes$ is the outer product \cite{novotny2012principles} (note $\varepsilon_0$ in the expression for polarizability, as required in SI units).
Combining this with the volume density of the two-level systems $\rho$ and assuming random orientation of transition dipole moments, one easily obtains the expression for the polarization density and permittivity of the medium:
\begin{equation} 
    \varepsilon = \varepsilon_{\infty} + f \frac{\omega_P^2}{\omega_0^2-\omega^2 - i\gamma \omega},
\end{equation}
where $\varepsilon_\infty$ is the non-resonant background permittivity, $\omega_P^2=\rho e^2/(3\varepsilon_0 m)$ is the plasma frequency, and the factor $1/3$ accounts for random isotropic orientation of dipoles in the medium.

\subsection{\label{sec:methods:slab} Eigenfrequencies of a slab}
Eigenfrequencies of a planar slab of thickness $L$ are found as poles of the reflection coefficient in the complex frequency plane. The reflection coefficient of a TE or TM polarized plane wave with wavenumber $k_0=\omega/c$ and in-plane momentum $k_x$ reads
\begin{equation}
    R_\mathrm{TE,TM}=\frac{r_\mathrm{TE,TM} (1-e^{2i k_{z,2} L})}{1-r_{\rm TE, TM}^2 e^{2i k_{z,3} L}}
\end{equation}
where 
\begin{equation}
    r_{\rm TE}=\frac{k_{z,1} - k_{z,2} }{k_{z,1} + k_{z,2}},~~
    r_{\rm TM}=\frac{k_{z,1} - k_{z,2}/\varepsilon(\omega) }{k_{z,1} + k_{z,2}/\varepsilon(\omega)}
\end{equation}
with $k_{z,1}=\sqrt{k_0^2 - k_x^2}$, $k_{z,2}=\sqrt{\varepsilon(\omega)k_0^2 - k_x^2}$ being the z-components of the wave vector in vacuum and dielectric, respectively.

$k_{z,1}$ has a branching point at $k_x=k_0$, and a cut needs to be made.
The field of the eigenmode outside the slab is described by the phase factor $e^{\pm i k_{z,1} z}$, where the plus and minus sign is applied in the regions $z>L/2$ and $z<-L/2$, correspondingly. The radiating modes above the light-line should have $\mathrm{Re}(k_{z,1})>0$, while proper waveguide modes have $\mathrm{Im}(k_{z,1})>0$.
Therefore, we make the branch cut along the negative imaginary axis through $\omega=-i \infty$ and choose the Riemann sheet of the square root function such that $\mathrm{Re} k_{z,1}>0$ above the light-line, and $\mathrm{Im}(k_{z,1})>0$ below the light-line. This choice ensures that we find radiating eigenmodes and proper (localized) waveguide modes. Choosing another Riemann sheet of the square root in the definition of $k_{z,1}$ results in dispersion of improper (diverging) modes below the light-line. The choice of sheet for $k_{z,2}$ is not important, because the slab contains both waves with $\pm k_{z,2}$.

\subsection{Dispersion of the TE$_1$ waveguide mode in a thin dielectric film}
Dispersion of TE waveguide modes in a dielectric film is given by the roots of the characteristic equation $1-r_{\rm TE}^2 e^{2i k_{z,2}L} =0$. This is a transcendental equation and in a general case cannot be solved analytically. However, in the case of an electrically thin film ($k_0 L \sqrt{\varepsilon_{\infty}}\ll 1$) the equation can be linearized and an approximate solution of the fundamental TE$_1$ mode can be found. 

For an electrically thin film, dispersion of the TE$_1$ mode is sandwiched between the light-line of the environment ($k_0=\omega/c$) and that of the slab ($k=\sqrt{\varepsilon_{\infty}}\omega/c$), and follows very close to the environment light-line; therefore, we can write the propagation constant of the guided mode as $k_x = k_0 + \delta k$, where $\delta k \ll k_0$.
The z components of the wave vectors in the environment and in the slab can be expanded as follows:
\begin{align*}
k_{z,1} &= \sqrt{k_0^2 - k_x^2} \approx i\sqrt{2k_0 \delta k}, \\ 
k_{z,2} &= \sqrt{\varepsilon k_0^2 - k_x^2} \approx k_0 \sqrt{\varepsilon-1}\left(1-\frac{\delta k}{(\varepsilon-1) k_0}\right).
\end{align*}
Note that $k_{z,1}$ scales as $O(\sqrt{\delta k})$, whereas the leading correction to $k_{z,2}$ scales as $O(\delta k)$.
The exponential phase factor can be expanded in a Tailor series $e^{2i k_{z,2}L} \approx 1+2i k_{z,2} L$. Plugging these expansions into the characteristic equation, we obtain:
\begin{align*}
    \left(\frac{k_{z,1} - k_{z,2} }{k_{z,1} + k_{z,2} }\right)^2 (1+2i k_{z,2} L)-1=0, \\
    i(k_{z,1}^2 - 2 k_{z,1} k_{z,2} + k_{z,2}^2)L=2k_{z,1}.
\end{align*}
Keeping only the leading power $O(\sqrt{\delta k})$ of the $k_x$ expansion and the constant term $O(1)$ in this equation, we find $\delta k = k_0^3 L^2 (\varepsilon-1)/8$, and the approximate dispersion relation of the TE$_1$, correspondingly, can be written as:
\begin{equation}
    k_x \approx k_0 + k_0^3 L^2 \frac{\varepsilon-1}{8}.
\end{equation}

\subsection{Eigenfrequencies of a cylinder}
TE (TM) modes of an infinitely long cylinder are defined as those having their electric (magnetic) field strictly perpendicular to the cylinder axis. Eigenfrequencies of TE01 and TM01 (monopole transverse electric and transverse magnetic) waveguide modes of an infinitely long circular cylinder of radius $a$ are found as roots of the characteristic equation.\cite{ishimaru2017electromagnetic,Doost2013}
\begin{align}
   \frac{\varepsilon}{k_{\rho,2}}  \frac{J_1'(k_{\rho,2}a) }{ J_1(k_{\rho,2}a) } -
      \frac{1}{k_{\rho,1}} 
      \frac{H_1'(k_{\rho,1}a) }{ H_1(k_{\rho,1}a) }&=0 ~~\rm{(TM)} \\
   \frac{1}{k_{\rho,2}}  \frac{J_1'(k_{\rho,2}a) }{ J_1(k_{\rho,2}a) } -
      \frac{1}{k_{\rho,1}} 
      \frac{H_1'(k_{\rho,1}a) }{ H_1(k_{\rho,1}a) }&=0 ~~\rm{(TE)}
\end{align}
where $J_1(z)$ and $H_1(z)=H_1^{(1)}(z)$ are Bessel and Hankel functions of the first kind, respectively, $k_{\rho,1}=\sqrt{k_0^2 - k_x^2}$ and $k_{\rho,2}=\sqrt{\varepsilon(\omega)k_0^2 - k_x^2}$ are the radial wavenumber of the eigenmode in vacuum and dielectric, respectively.

The definition of $k_{\rho,1}$ requires a branch cut at $k_{\rho,1}=k_0$.
Radial dependence of the eigenmode field outside the waveguide is given by the factor $H_1(k_{\rho,1}\rho)$. Therefore, making the branch cut going along the negative imaginary axis through $\omega=-i \infty$ and choosing the Riemann sheet yielding $\mathrm{Re}(k_{\rho,1})>0$ above the light-line results in the spectrum of radiating eigenmodes and localized waveguide modes.\cite{Doost2013}

Eigenfrequencies of the HE11 (dipole) mode are found as roots of the following characteristic equation \cite{ishimaru2017electromagnetic}
\begin{multline}
\left[\frac{1}{k_{\rho,2}}  \frac{J_1'(k_{\rho,2}a) }{ J_1(k_{\rho,2}a) } -
   \frac{1}{k_{\rho,1}} 
   \frac{H_1'(k_{\rho,1}a) }{ H_1(k_{\rho,1}a) }\right]
   \left[\frac{1}{k_{\rho,2}}  \frac{J_1'(k_{\rho,2}a) }{ J_1(k_{\rho,2}a) } - \right.\\
      \left.\frac{1}{k_{\rho,1}} 
   \frac{H_1'(k_{\rho,1}a) }{ H_1(k_{\rho,1}a) }\right] = 
   \left[\frac{k_x}{k_0}
   \frac{(\varepsilon-1)k_0^2 }{a^2 k_{\rho,1}^2 k_{\rho,2}^2 }\right]^2 ~~\rm{(HE11)}
\end{multline}

\subsection{Eigenfrequencies of a sphere}
Eigenfrequencies of Mie modes with orbital number $l$ of a sphere of radius $a$ and permittivity $\varepsilon=n^2$ in vacuum are found as the roots of the characteristic equation \cite{Bohren2004}
\begin{align}
    \psi_l(n x) \xi_l '(x)- n \xi_l(x) \psi_l '(nx)&=0 ~~ \rm{(TE)}\\
    n\psi_l(n x) \xi_l '(x)- \xi_l(x) \psi_l '(nx)&=0 ~~ \rm{(TM)}
\end{align}
where $x=k_0 a$, $\psi_l(x)=x j_l(x)$ and $\xi_l(x)=x h_l^{(1)}(x)$ are Ricatti-Bessel functions, and $j_l(x)$ and $h_l^{(1)}(x)$ are spherical Bessel and Hankel functions of the first kind, respectively.

\begin{acknowledgments}
A.C., D.G.B and T.S. acknowledge the financial support from the Swedish Research Council (VR project grant No: 2017-04545 and VR Miljö grant No: 2016-06059). T.J.A. acknowledges support from the Polish National Science Center (project 2019/34/E/ST3/00359).
\end{acknowledgments}

\nocite{*}
\bibliography{biblio} % Produces the bibliography via BibTeX.

%%%%%%%%%%%%%%%%%%%%%%%%%%%%%%%%%%%%%%%%%%%%%%%%%%%%%%%%%

\end{document}

\bibliographystyle{apsrev} 

\title{Supporting Information \\ Abundance of cavity-free polaritonic states in resonant materials and nanostructures} 

\author{Adriana Canales}
\affiliation{Department of Physics, Chalmers University of Technology, 412 96, Göteborg, Sweden.
}%

\author{Denis G. Baranov}%
\affiliation{Department of Physics, Chalmers University of Technology, 412 96, Göteborg, Sweden.
}%
\affiliation{Center for Photonics and 2D Materials, Moscow Institute of Physics and Technology, Dolgoprudny 141700, Russia.
}

\author{Tomasz J. Antosiewicz}
\affiliation{Faculty of Physics, University of Warsaw, Pasteura 5, 02-093, Warsaw, Poland.
}%
\affiliation{Department of Physics, Chalmers University of Technology, 412 96, Göteborg, Sweden.
}

\author{Timur Shegai}%
 \email{timurs@chalmers.se.}
\affiliation{Department of Physics, Chalmers University of Technology, 412 96, Göteborg, Sweden.
}%

\maketitle

%\begin{article}

\section{Bulk Polaritons}

\begin{figure}[h!]
  \includegraphics[width=\linewidth]{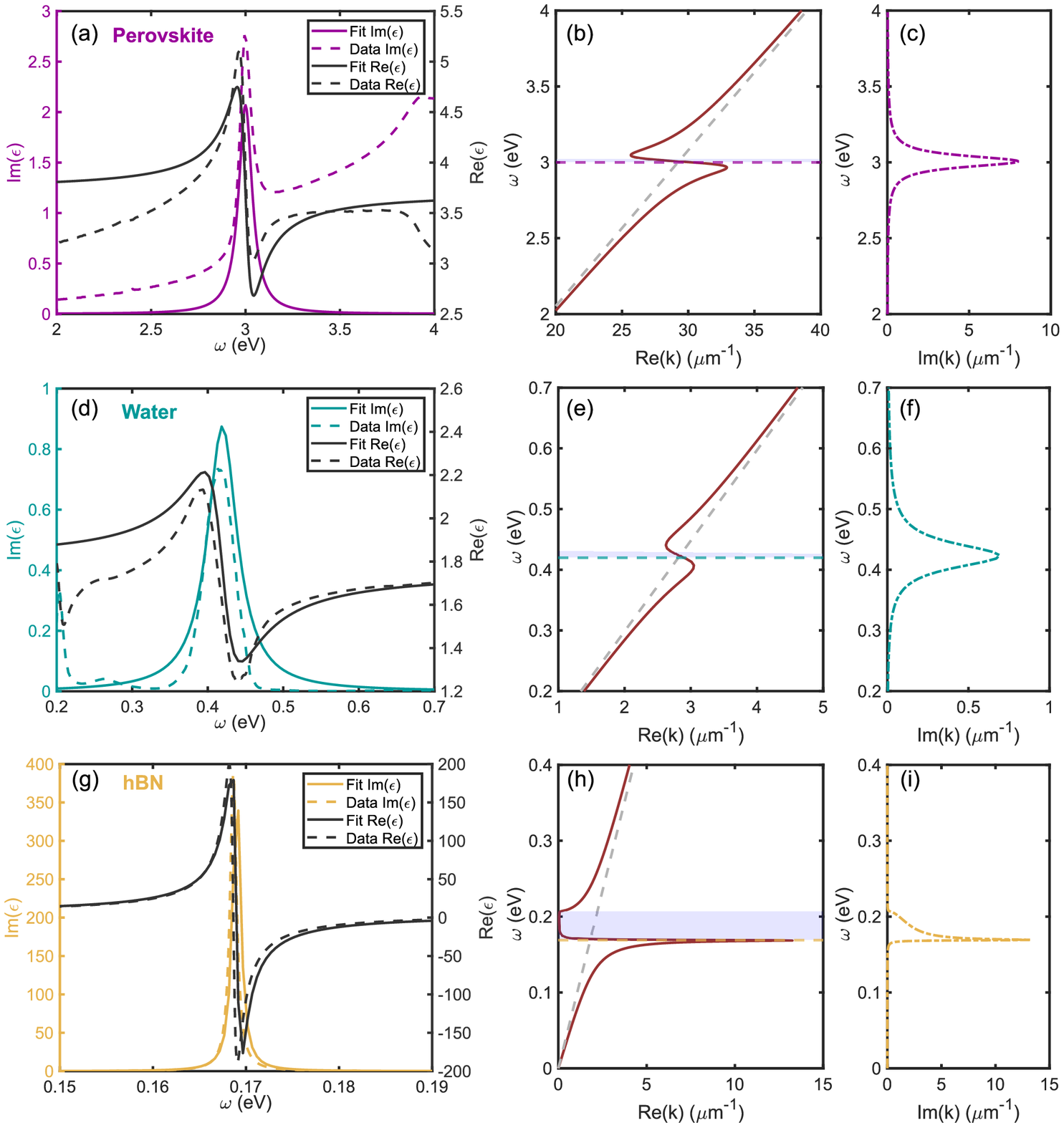}
  \caption{Permittivities and bulk polariton dispersion with complex$-k$. The first column shows the data (dashed lines) of permitivitties and the Lorentzian fitting (solid lines) used in the calculations (a) CsPbCl$_3$ perovskite \cite{tiguntseva2018tunable} (b) water \cite{segelstein1981complex} and (g) hBN \cite{Caldwell2015}. The second column shows the bulk polariton dispersion with $\operatorname{Re}(k)$  of each material (b,e,h). The light-line and the uncoupled exciton are marked in dashed lines. Finally, the third column (c,f,i) shows the damping of such polaritons given by the   $\operatorname{Im}(k)$ vs $\omega$ in the same order as mentioned previously: perovskite, water and hBN. }
  \label{SIfig:BULK}
\end{figure}

In this section we explain how we obtain the poles in the complex$-\omega$ plane, we show the fittings for the permittivities used throughout the text, the bulk polariton dispersion with the complex$-k$ and the values of the bulk Rabi splitting, $2g_0$, for all the materials used.

First, the video in the supporting material shows the poles in the complex$-\omega$ plane of a bulk TDBC J-aggregate  with a permittivity given by $\omega_0=2.11$ eV, $\gamma=0.1$ eV, $\varepsilon_{\infty}=2.15$ and $f \omega_P^2 =0.445 ~\text{eV}^2$, \cite{balasubrahmaniyam2020coupling} for different values of $k$ from 5 to 25 $\mu m^{-1}$. 
The position of the poles, for each $k$ gives us the information of both dispersion --$k$ vs $\operatorname{Re}(\omega$-- and trajectories of the poles in the complex-$\omega$ plane -- $\operatorname{Re}(\omega)$ vs $\operatorname{Im}(\omega)$ -- of the eigenmodes shown in Fig. 2(b,e). The value of the intensity is not considered in the main text, but it relates to the excitation rate of the specific mode. The video shows a clear Rabi splitting at $k=16 \mu m^{-1}$ and depicts how the lower and the upper polariton are separated by the polariton gap, which can be found as defined in equation 6 in the main text.  

The first column of Fig. S\ref{SIfig:BULK} shows, in dashed lines, the data of the permittivity of (a) perovskites  CsPbCl$_3$, (d) water and (g) hBN. As well as, the Lorentz model fitting for each material, in solid lines, which was used for the calculations. Note that all the frequencies $\omega$ in the labels correspond to $\hbar \omega$ since the units used throughout the paper are eV.

The second column of Fig. S\ref{SIfig:BULK} shows the bulk polariton dispersion by considering complex$-k$  ($k=\sqrt{\varepsilon(\omega)}\omega/c$) for hBN, perovskites and water.The same information for J-aggregates is found in the upper panels of Fig. 2 in the main text. We can also see the two branches of each bulk polariton (b,e,h) connected by a region of anomalous dispersion. 
Finally, the last column shows the damping of the solutions within the polariton gap due to the large values of $\operatorname{Im}(k)$. 

\begin{table}
\centering
\caption{\label{SItab:tableBulk} The table shows the calculated values of the bulk Rabi splitting, $2g_0$, for the materials analyzed in this study.}
\begin{tabular}{||c | c||}
\hline
\textbf{Material} & \textbf{$2g_0$ (meV)} \\
\hline
\hline
Water & 100 \\
hBN & 120\\
Perovskite (CsPbCl$_3$) & 382 \\
J-aggregates low $f$ & 455 \\
J-aggregates high $f$ & 1204 \\
\hline
\end{tabular}
\end{table}

%%%%%%%%%%%%%%%%%%%%%% 2D Slabs %%%%%%%%%%
\newpage

\section{2D: Polaritons in Lorentz slabs}
\begin{figure*}[t]
  \includegraphics[width=\linewidth]{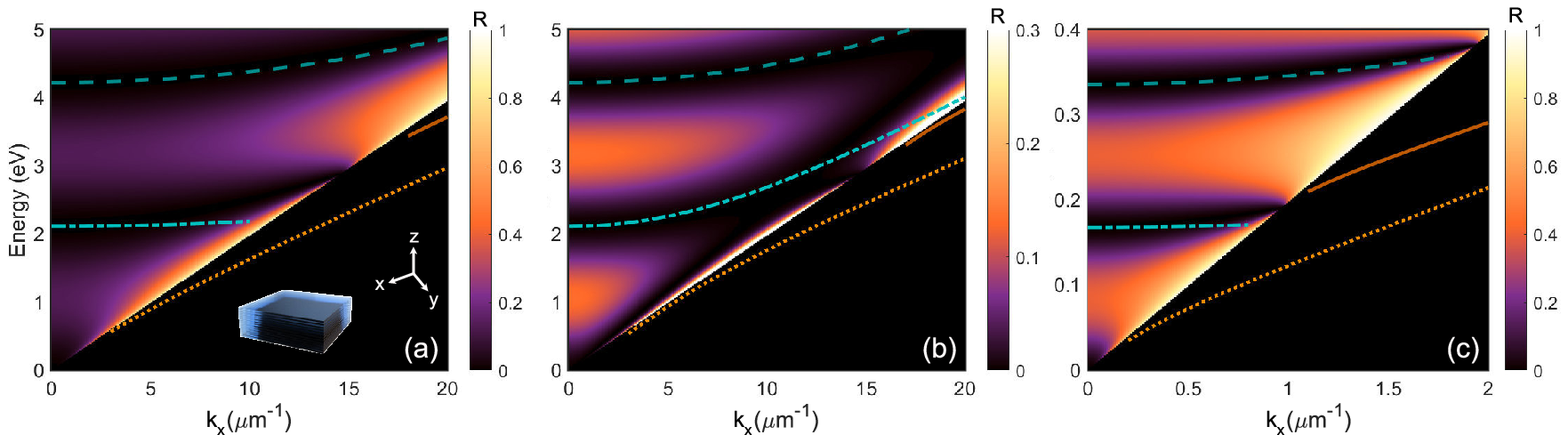}
  \caption{Reflectivity (colormap) and modes of the cavities without a resonance ($f=0$). The modes above the light-line are radiating FP modes (blue dot-dashed and dashed lines) and the modes below the light-line are waveguide modes (orange dotted and solid lines) for (a) TE polarization modes and (b) TM polarization of a 200 nm thick slab with the J-aggregates permittivity. (c) TE polarization of a 1.75 $\mu$m thick slab of hBN.}
  \label{SIfig:2D_bare}
\end{figure*}

This section shows the slabs calculations for the TE trajectories of the radiating modes, the uncoupled optical modes, poles in the complex$-\omega$ plane and the calculations for hybridization with TM modes.

Figure S\ref{SIfig:2D_bare} shows the dispersion of the uncoupled optical modes above (in blue) and below (in orange) the light-line (obtained by setting $f=0$). The poles are plotted on top of the colormap for the reflectivity of the slab. The first two panels correspond to a 200 nm thick J-aggregate slab with (a) for TE polarization and (b) for TM. (c) shows the same information for the TE modes of a 1.75 $\mu m$ hBN slab. Such information is necessary to find  $k_x$ such that there is zero detuning between the resonance and the optical mode.

\begin{figure}[b]
  \includegraphics[width=\linewidth]{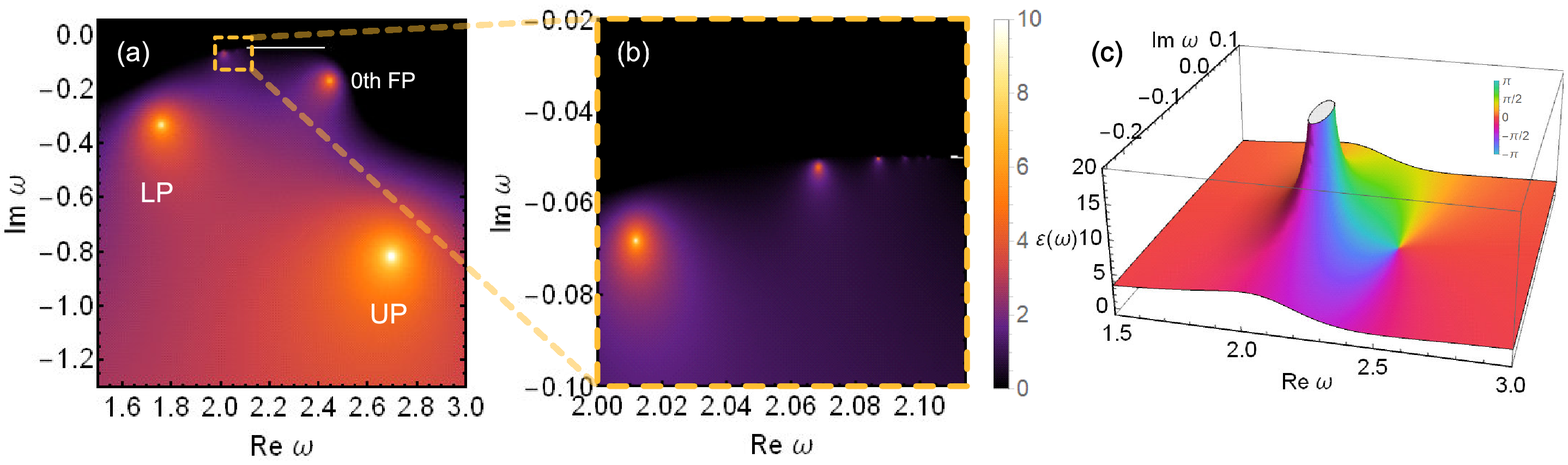}
  \caption{(a) Poles in the complex$-\omega$ plane of a 200 nm thick J-aggregate slab with TE polarization with $f \omega_P^2 =3.116 ~\text{eV}^2$ at $k_x=0$ (radiative modes). (b) Close-up on the countable poles in the vicinity of $\omega_0-i\gamma /2=2.11 - i0.05$. (c) Permittivity of the same J-aggregates in the complex$-\omega$ plane. }
  \label{SIfig:SI_poles}
\end{figure}

Figure S\ref{SIfig:SI_poles} shows the poles in the complex$-\omega$ plane (similar to the ones in the supporting video) for a  200 nm thick J-aggregate slab $f \omega_P^2 =3.116 ~\text{eV}^2$ at normal incidence, $k_x=0$, thus all of the poles are radiating modes corresponding to the upper polariton (UP), the lower polariton (LP) and the 0$^{th}$-order FP mode below the exciton gap. The countable many poles in the vicinity of $\omega_0-i \gamma/2$, are described carefully in the main text. In short, they correspond to higher order modes given by the steep increase of $\varepsilon(\omega)$, due to the pole of the permittivity, as shown in \ref{SIfig:SI_poles}(c). It is important to note that in the dispersion and trajectories plots we only show the lower and upper polariton trajectories, no other modes are shown for the sake of clarity, but they also exist.

\begin{figure}[t]
  \includegraphics[width=\linewidth]{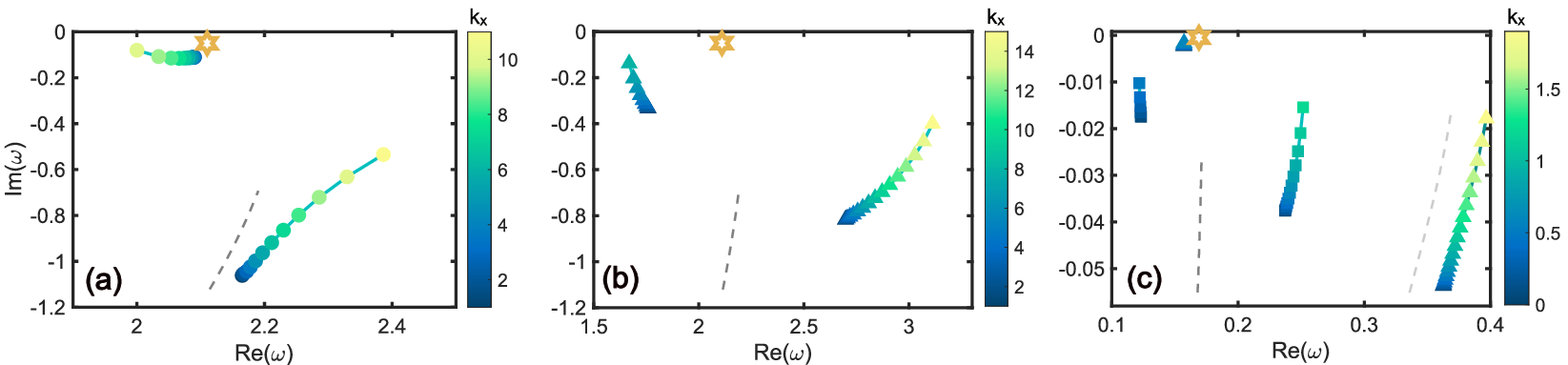}
  \caption{Trajectories of the poles of the hybridized TE radiating Fabry-Pérot (FP) modes in the slab for (a)  J-aggregates with low f, (b) J-aggregates with high f and (c) 1.75 $\mu$m thick slab of hBN. In (c) $\square$ corresponds to the 1$^{st}$ FP mode and $\vartriangle$ to the 2$^{nd}$ one. The colormap represents the value of $k_x$ in $\mu m^{-1}$.}
  \label{SIfig:2D_TE_traj}
\end{figure}

Figure 3 in the main text shows the dispersion of the eigenmodes and their trajectories of the ones resulting from the hybridization with TE waveguide modes. Here, fig. S\ref{SIfig:2D_TE_traj} shows the trajectories in the complex-$\omega$ plane of the hybridized modes with the radiating Fabry-Pérot (FP) modes of TE polarization. We show the trajectories for two J-aggregates, (a) with low f and (b) with high f, and for hBN. The uncoupled modes are shown in grey dashed lines and the location of the excitonic resonance is shown in a yellow star. These hybridized modes are much harder to follow, in comparison with the waveguide modes because the uncoupled optical mode has a non-trivial trajectory (showed in dashed lines) with high losses ($\operatorname{Im}(\omega)$. Thus, when it is coupled to the exciton it is attracted by it with a non-trivial trajectory. An additional issue with this modes is the cut-off induced by the light-line, such that after a certain $k_x$ the modes become guided.

\begin{figure}[h]
  \includegraphics[width=\linewidth]{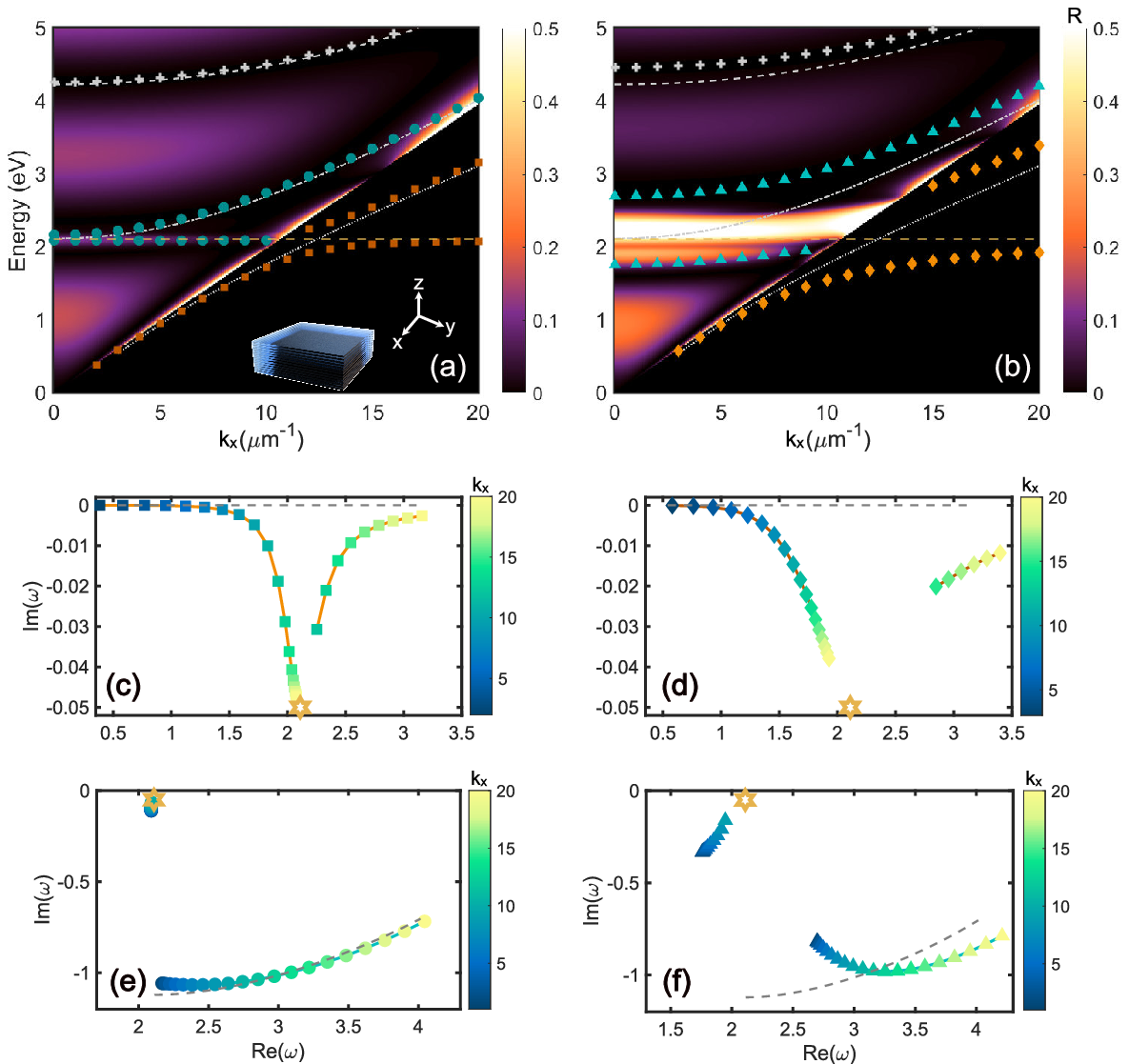}
  \caption{Reflectivity and dispersion of the poles of the self-hybridization with TM modes of a 200 nm thick J-aggregate slab with (a) $f \omega_P^2 =0.445 ~\text{eV}^2$ (b) with $f \omega_P^2 =3.116 ~\text{eV}^2$. The trajectories of the poles in the complex-$\omega$ plane for the low f case are shown in (c) for the 1$^{st}$ guided mode -- TM$_1$ -- and (e) for the 1$^{st}$ FP mode. The same information for the high f case is in (d) for TM$_1$ and (f) for the 1st FP mode. The value of $k_x$ is shown in the color of the marker.}
  \label{SIfig:2DTMJagg}
\end{figure}

Fig. S\ref{SIfig:2DTMJagg} depicts the analogue information as in the Fig. 3 in the main text, but for TM modes. In this case, the dispersion and the trajectories of the poles are given by the self-hybridization of the J-aggregate slab with the TM modes both below and above the light line. 
The 1$^{st}$ FP mode is so weakly coupled that the trajectories barely change from the uncoupled ones (Fig. S\ref{SIfig:2DTMJagg}(e)), as expected there is no splitting noticeable in the dispersion (Fig. S\ref{SIfig:2DTMJagg}(a)). Nevertheless, TM$_1$ is strongly coupled, the coupling strength is part of the summary shown in Fig. 6 in the main text. 
On the other hand, when $f \omega_P^2 =3.116 ~\text{eV}^2$, there is splittig in the FP mode, thus the optical mode is clearly pulled towards the exciton in the trajectories plot (S\ref{SIfig:2DTMJagg}(f)). The splitting with TM$_1$ is so big that the mode at zero detuning for the upper polariton is cut by the light-line.

\newpage

\section{1D: Long cylinder}

\begin{figure}[h]
  \includegraphics[width=\linewidth]{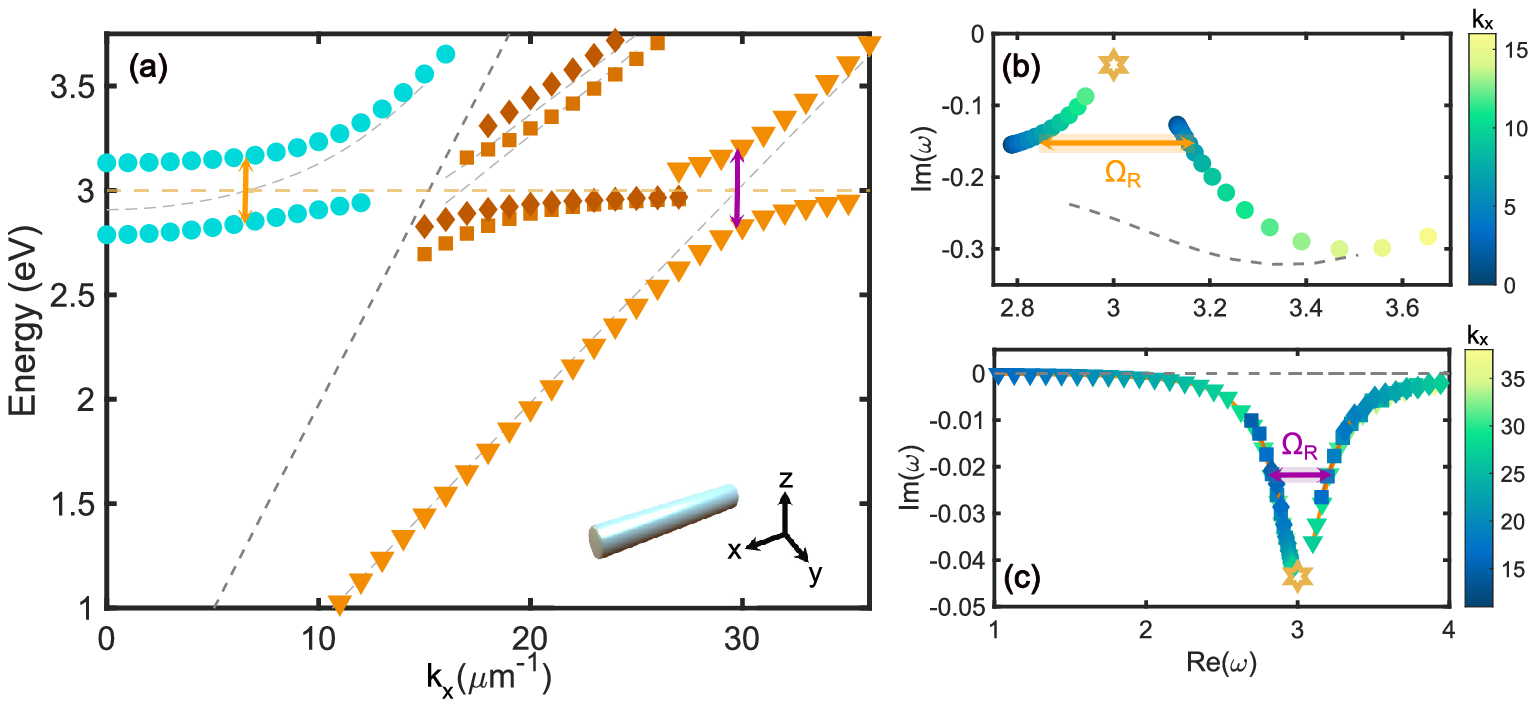}
  \caption{\emph{\textbf{Polaritons in inifinite cylinders with TEM modes}}. (a) Dispersion of the polaritons given by the hybridization of the resonance in the perovskite CsPbCl$_3$ with TEM polarized radiating modes (4$^{th}$ mode, $\circ$) and the guided modes where HE11 mode, $\triangledown$, is the only one without cutoff. 5$^{th}$ in $\square$ and 6$^{th}$ in $\Diamond$). The trajectories of such poles in the complex-$\omega$ plane are shown in (b) for the radiating and (c) the guided modes. The colorbar shows the value of $k_x$ and the colored arrows show the Rabi splitting above (orange) and below (purple) the light-line. The uncoupled optical modes are shown in dashed lines.}
  \label{SIfig:1DTEM}
\end{figure}

This section contains the calculations corresponding to infinite perovskite CsPbCl$_3$ cylinders with $r=250$ nm, but showing the self-hybridization with the TEM and TE modes, in analogy to in Figure 4 in the main text. The figures do not show all the modes in the plotted energy spectra. The only radiating mode for TEM in Fig. S\ref{SIfig:1DTEM} the plot is the 4$^{th}$ mode, which shows the Rabi splitting. TEM has many guided modes that couple to the resonance, the only modes shown in the plot are the HE11 mode (the first guided one) and the 5th and 6th modes, which are the last modes crossing the resonance, $\omega_0 = 3$ eV due to the light-line cut. As we can see in Fig. S\ref{SIfig:1DTEM}(c), all the modes below the light-line follow the same trajectories but at different values of $k_x$. The Rabi splitting above the light-line is shown in an orange arrow and below it with a purple arrow. Both values of the Rabi splitting were considered for the summary plot in Fig. 6(b) in the main text.

\begin{figure}[h]
  \includegraphics[width=\linewidth]{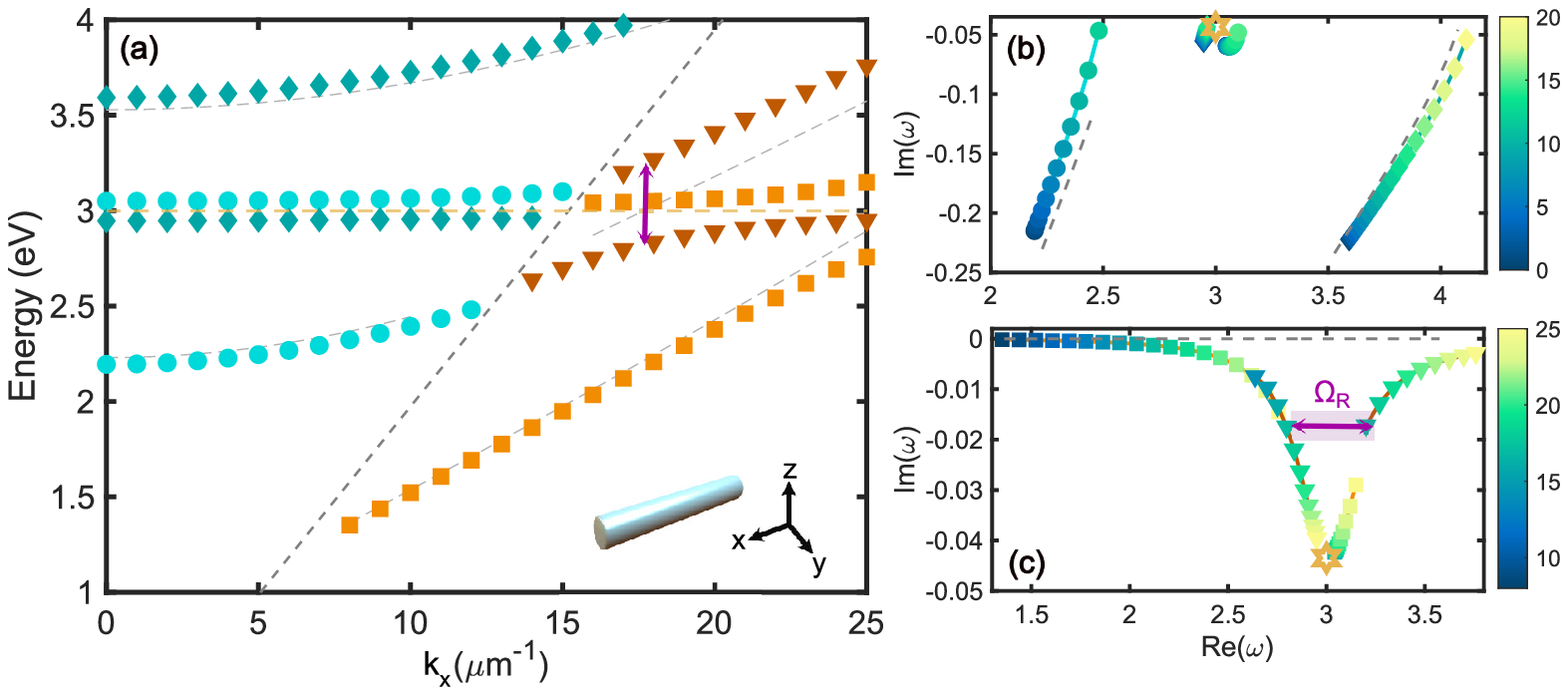}
  \caption{\emph{\textbf{Polaritons in inifinite cylinders with TE modes}} Polaritons in an infinite CsPbCl$_3$ cylinder with a radius of 250 nm hybridized its TE optical modes. (a) Shows the dispersion of the poles, with the 2$^{nd}$ radiating mode $\circ$, the 3$^{rd}$ one in $\Diamond$ and the 1$^{st}$ guided mode TE$_{01}$, $\square$, and the 2$^{nd}$ TE$_{02}$, $\triangledown$. The trajectories of the poles in the complex-$\omega$ plane of the modes (b) above and  (c) below the light line. The Rabi splitting is shown with a purple arrow.}
  \label{SIfig:1DTE}
\end{figure}

Fig. S\ref{SIfig:1DTE} shows the information (same material and dimensions) but for TE modes. In this case the radiative modes closest to the resonance are the 2$^{nd}$ ($\circ$) and 3$^{rd}$ ($\Diamond$) modes. Both modes are highly detuned, thus we cannot see the Rabi splitting above the light-line for this radius. The detuning is also visible in the trajectories (Fig. S\ref{SIfig:1DTE}(b)), where the modes are barely modified with respect to the uncoupled ones (dashed lines and star).
On the other hand, below the light-line there is a clear Rabi splitting marked with a purple arrow for both TE$_{01}$, $\square$ , and TE$_{02}$, $\triangledown$. 

\newpage

\section{0D: Polaritons in spheres}

This section shows the dispersion and the trajectories of the TM$_{1,1}$ mode of water spheres. Since the loss of the optical mode is too large (in comparison to the resonance loss $\gamma$), the optical mode and the resonance are only weakly coupled. This causes an crossing in the dispersion $\operatorname{Re}(\omega)$ vs radius (a similar behavior was observed in Fig. 2(a)). On the other hand, in the complex-$\omega$ plane we observe just a small deviation of the trajectory of the optical mode and a very small circular trajectory around the $\omega_0 - i\gamma/2$ (similar to the trajectory in Fig.2(d)). The dispersion in the case of spheres is given by different sizes of spheres, thus the colormap in this case shows the radius of the sphere in $\mu m$.

\begin{figure}[h] %%%%Water%%%%
  \includegraphics[width=\linewidth]{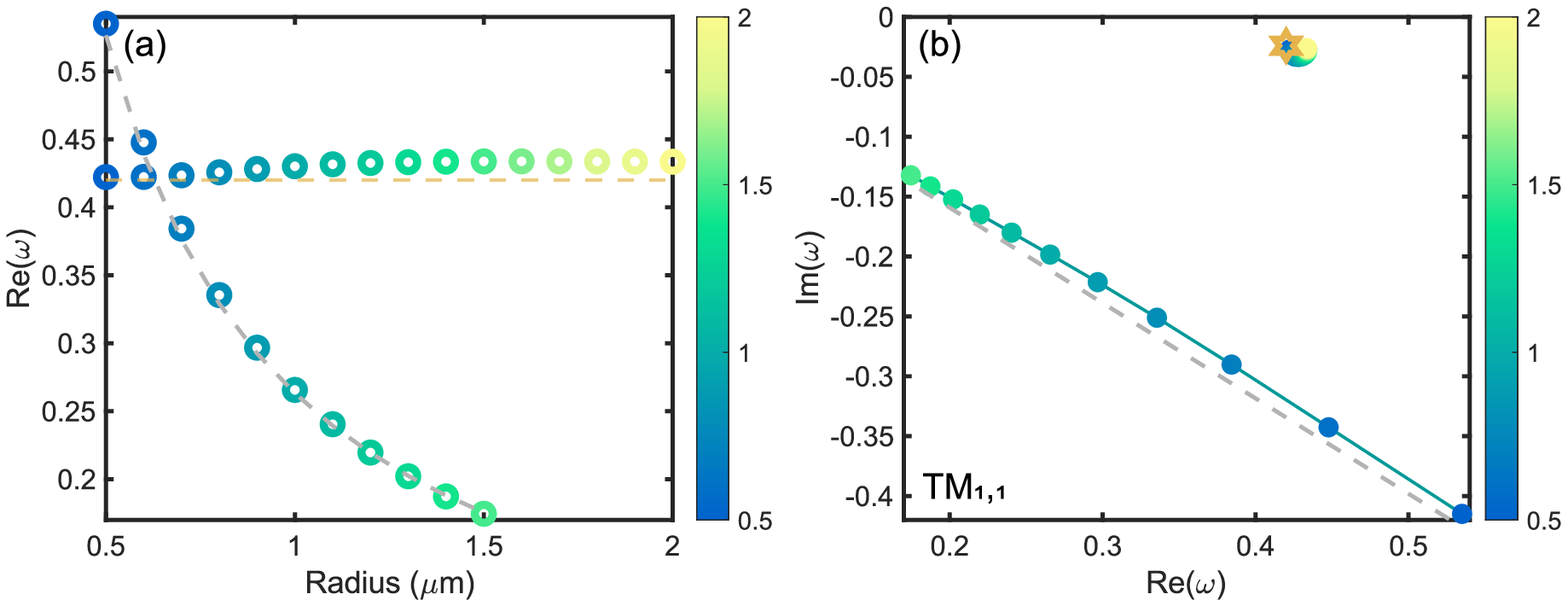}
  \caption{Weak coupling in water spheres TM$_{1,1}$. (a) Eigenmodes dependence on the radius of the spheres. (b) Trajectories of the poles in the complex$-\omega$ plane for different radius. The colormap shows the value of the radius in $\mu m$.}
  \label{figSI:1DTEM}
\end{figure}

\newpage
\section{Limit of Rabi splitting}

Table \ref{SItab:tableBulk} contains the calculated values of the bulk Rabi splitting, $2g_0~=~\omega_P\sqrt{f /\varepsilon_\infty}$, for the studied permittivities. Such values were used to normalize the obtained Rabi splitting of the different structures of Fig. 6 in the main text. Due to the difference in materials, the structures (spheres, cylinders and slabs) have different dimensions. 

For the 2D case, the Rabi splitting for perovskites was obtained for slabs with a thickness of  $L=106$ nm; for water we used $L = 1.1~\mu$m; and as mentioned in the text before, for J-aggregates we used  $L = 200$ nm and for hBN $L=1.75~\mu$m.

For the 1D case, the Rabi splitting for water was obtained from cylinders with a radius of $r=2~\mu$m; for J-aggregates we used  $r = 400 $ nm; and for pervoskites we have mentioned already that $r=250 $ nm.

For the 0D case, spheres of different radius were used for obtaining the Rabi splitting of different modes. The smallest spheres are coupled to TE$_{1,1}$. Thus this is the minimal radius for spheres sustain polaritons. In the case of perovskites that radius was $r = 98$ nm; for J-agregates it was $r=180$ nm and for water we have mentioned before that $r= 1~\mu$m.

\bibliography{biblio}